\documentclass[aps,twocolumn,nofootinbib,showpacs,floatfix]{revtex4}
\usepackage{graphics} 
\usepackage{amssymb,amsmath}
\usepackage{amsmath}
\usepackage{psfrag}
\usepackage{graphicx}
\usepackage{subfig}
\usepackage{multirow}
\usepackage{color}
\usepackage{layout}
\usepackage{textcomp}
\usepackage{hyperref}
\usepackage{ulem}

\def\ep{\epsilon}
\def\ga{\gamma}

\def\be{\begin{equation}} 
\def\ee{\end{equation}} 
\def\bea{\begin{eqnarray}} 
\def\eea{\end{eqnarray}}  
\def\bean{\begin{eqnarray*}} 
\def\eean{\end{eqnarray*}} 
\def\dd{\partial}  
 
\def\tv{{\text v}}

\def\bv{{\bf v}}
\def\br{{\bf r}}

\def\bse{\begin{subequations}}
\def\ese{\end{subequations}}

\def\bF{{\mathbf F}}

\def\bV{{\mathbf V}}

\def\lsim{\raise 0.4ex\hbox{$<$}\kern -0.8em\lower 0.62ex\hbox{$\sim$}} 
\def\gsim{\raise 0.4ex\hbox{$>$}\kern -0.7em\lower 0.62ex\hbox{$\sim$}}

\def\f0N{f_0^{(N)}}
\def\bec{\begin{center}}
\def\eec{\end{center}}

\begin{document} 

\title{{\bf Formation and relaxation of quasi-stationary states 
in particle systems with power law interactions}}

\author{B.~Marcos$^{1,2}$, A. Gabrielli$^{3,4,5}$ and M. Joyce$^{6}$} 
\affiliation{$^1$Universit\'e C\^ote d'Azur, CNRS UMR 7351, LJAD, France\footnote{Parc Valrose 06108 Nice Cedex 02, France}} 
\affiliation{$^2$Instituto de F\'{\i}sica, Universidade Federal do Rio Grande do Sul, Brazil\footnote{ Caixa Postal 15051, CEP 91501-970, Porto Alegre, RS, Brazil}}
\affiliation{$^3$Istituto dei Sistemi Complessi (ISC) - CNR, Via dei Taurini 19, 00185-Rome, Italy}
\affiliation{$^4$IMT - Institute of Advanced Studies,  Piazza S. Ponziano, 6, 55100-Lucca, Italy}
\affiliation{$^5$London Institute for Mathematical Sciences (LIMS), South Street 22, Mayfair London, UK}
\affiliation{$^6$Laboratoire de Physique  Nucl\'eaire et de Hautes \'Energies, UPMC IN2P3 CNRS  UMR 7585, Sorbonne Universit\'es, 4, place Jussieu, 75252 Paris Cedex 05, France}

\begin{abstract}   
\begin{center}    
{\large\bf Abstract}
\end{center}    
We explore the formation and relaxation of so-called  quasi-stationary states (QSS) for 
particle  distributions in three dimensions  interacting via  an attractive radial pair potential  $V(r \rightarrow \infty) \sim 1/r^\gamma$  with $\gamma > 0$, 
and either a soft-core or hard-core regularization at small $r$. 
In the first part of the paper we generalize, for any spatial dimension 
$d \geq 2$, Chandrasekhar's approach for the case of gravity to obtain 
analytic estimates of the rate of collisional relaxation due to two 
body collisions.
The resultant relaxation rates indicate an essential 
qualitative difference depending on the integrability of the pair force at large distances:  
for $\gamma >d-1$ the rate diverges in the large
particle number $N$ (mean field) limit, unless a 
sufficiently large soft core is present; for $\gamma < d-1$, on the 
other hand, the rate vanishes in the same limit even
in the absence of any regularization. In the second part of the paper we compare our analytical predictions 
with the results of extensive parallel numerical simulations in $d=3$ performed 
with an appropriate modification of the 
GADGET code, for a range of different exponents $\gamma$ and 
soft cores leading to the formation of QSS.  We find, just as for the 
previously well studied case of gravity (which we also revisit), 
excellent agreement between the parametric dependence of the
observed relaxation times and our analytic predictions. 
Further, as in the case of gravity, we find that the results
indicate that,  when large impact factors dominate,
the appropriate cut-off is the size of the system (rather than, 
for example, the mean inter-particle distance). 
Our results provide strong evidence that the existence of
QSS is robust only for long-range interactions with
a large distance behavior $\gamma < d-1$; for 
$\gamma \geq d-1$ the existence of such states will be 
conditioned strongly on the  short range properties of 
the interaction.
\end{abstract}    
\maketitle   
\date{today}

\section{introduction}

 There are many systems of particles interacting with long-range
 interactions in nature: self-gravitating bodies in astrophysics and
 cosmology \cite{binney}, two-dimensional fluid dynamics
 \cite{chavanis_book}, cold atoms \cite{chalony_13}, etc. Considering,
 for simplicity, $d-$dimensional particle systems which interact
 through an isotropic pair potential $v(r)$, long-range systems are
 usually defined as those for which
\be
\label{pot}
v(r\to\infty)\sim\frac{g}{r^\gamma},
\ee
where $\gamma \leq d$, and $g$ is a coupling constant. This
characterization of interactions as long-range arises in equilibrium
statistical mechanics \cite{campa_etal_LRreview_2009}: in a system of $N$ particles 
in a volume $V$,  the average energy of a particle is, for $\gamma > d$, 
independent of the size of the system in the ``usual" thermodynamic limit $N
\rightarrow \infty$, $V \rightarrow \infty$ at fixed density
$N/V$. For $\gamma \leq d$ a different thermodynamic limit must 
be taken in order to recover extensivity of the thermodynamic 
potentials, and $N$ independent intensive properties of the system,
as $N \rightarrow \infty$.  More specifically, the potential energy
$\Phi_i$ of a particle scales as $\Phi_i \sim g N /V^{\gamma/d}$ and
$g$ and $V$ must be scaled appropriately with $N$ so that $\Phi_i$ is
constant.  This is usually called the mean-field thermodynamic limit
(or the Vlasov limit when is is taken at fixed system size).
Using this scaling, the
total energy becomes extensive and it is possible
to compute thermal equilibrium properties.
For the class of systems we consider here, with
attractive power law interactions  at large scales in three dimensions,
such a treatment has been given in  
\cite{ispolatov_cohen}. 
For $\gamma<d$ they present  unusual features compared to short range 
systems: inhomogeneous spatial distributions, inequivalence of
the statistical ensembles, negative specific heat in the 
microcanonical ensemble etc.\footnote{All these considerations 
are for classical systems. For studies of properties
of quantum spin systems with power law interactions see
e.g. \cite{Kastner_2011,Cevolani_2016}.}.

For the case of gravity it was understood decades ago, however, in the
context of astrophysics (through the seminal works of Chandrasekhar,
Lynden-Bell and others) that such considerations based on equilibrium
statistical mechanics are only relevant physically on time scales very
long compared to those on which such systems evolve dynamically
(e.g. the formation and evolution of galaxies), and that the scenario
of the dynamics of such systems is completely different to that of
short range systems: on a timescale $\tau_{dyn}$ characteristic of the
mean field dynamics (and independent of $N$ in the mean field limit
described above) one observes the formation, under the effect of a
mean field global interaction through so-called mean-field relaxation, of
very slowly evolving macroscopic states (e.g. galaxies) which are far
from thermal equilibrium.  For gravity in $d=3$ dimensions, the time
scale for evolution towards equilibrium was first estimated by
Chandrasekhar \cite{chandra_42} to be $\tau_{coll} \sim (N/\ln N)
\tau_{dyn}$. Thus as $N \rightarrow \infty$ in the mean field limit,
the system remains trapped in such states and never evolves towards
thermodynamic equilibrium.  A similar phenomenology has been
established in the last years in the study of various other systems
with long-range interactions (see e.g.
\cite{yamaguchi_etal_04,joyce_10,teles_10,marcos_13b}): 
relaxation on a mean field time scale to a ``quasi-stationary state"
(QSS) followed by a relaxation towards thermodynamic equilibrium on a
time scale which diverges with the particle number $N$.  This scenario
has thus been proposed as a kind of paradigm for the dynamics of this
class of interactions (e.g.~\cite{campa_etal_LRreview_2009,levin_14,chavanis_12c}).

More formally the evolution of a system of
$N$ particles interacting through the pair potential 
\eqref{pot} can be described by the 
equation
\be
\label{vlasov}
 \frac{\dd f}{\dd t}+\bv(\br,t)\cdot\frac{\dd f}{\dd
    \br}+\bF[f]\cdot\frac{\dd f}{\dd \bv}=C_N\,,
\ee
where $f(\br,\bv,t)$ is the  mean phase space density function,
i.e., the density of particles at the position $\br$ with
velocity $\bv$ at time $t$, and  $C_N$ is called the ``collision term".
In general the latter is a functional of the $n-point$ distribution 
functions. The term  $\bF[f]$ is the {\it mean field force} which 
can be written in terms of the pair potential $v(r)$ as
\be
\label{force-vlasov}
\bF[f]=-\int f(\br',\bv,t)\nabla_{\br} v(|\br-\br'|)d\br'd\bv\,.
\ee 
A mean-field dynamical description is valid if, in the 
mean-field (or Vlasov) limit, we have that  
\be
\label{coll_limit}
\lim_{N\to\infty} C_N=0
\ee
in which case the dynamics is described by the
Vlasov equation, known as the ``collisionless Boltzmann
equation" in the astrophysical literature (e.g. \cite{binney}). 
QSS are understood as stable stationary solutions of
these equations, and mean-field relaxation as the 
evolution towards such states in the same mean-field
framework (on timescales of order $\tau_{dyn}$).
Correspondingly, in any finite (but large) 
$N$ system, the term $C_N$ then describes the 
``collisional" corrections to the mean-field dynamics.

For long-range interactions, therefore, to show that QSS should exist
one should analyze these collision terms, and determine firstly that
they do indeed satisfy the condition (\ref{coll_limit}). Further  in order to
understand their evolution away from QSS at large but finite $N$, and (possibly) towards
thermal equilibrium, one needs to derive a suitable kinetic theory,
which should allow one to infer the scalings of the time scale (or
scales) characterizing such evolution  as a function of $N$. Concerning the first step
rigorous results have been obtained showing that the limit does exist
in the gravitational case \cite{braun+hepp,spohn} and for any
 potential with $\gamma\le 1$ \cite{hauray_07} (both
in $d=3$ dimensions) and provided a suitable regularization
(i.e. softening) of the potential is imposed at small separations
(see also \cite{Boers+pickl_2013,Lazarovici+pickl_2015}).
However these provide only rigorous lower bounds ($\sim log N$) to the
time scales on which the Vlasov dynamics is valid. They do not allow
us to calculate in any practical manner the time scales for
collisional relaxation, nor even to determine their parametric
scalings. Many attempts have been made in this direction through the
construction of explicit kinetic theories
\cite{weinberg_93,heyvaerts_10,chavanis_12b,fouvry_15a,
fouvry_15b,fouvry_15c,fouvry_16a,fouvry_16b,benetti_16} but, in
practice 
it is difficult to apply these methods to realistic systems to establish the relevant time scales, and in particular their parametric
scalings.
Moreover, these theories do not take into account strong collisions.  
Often (e.g. \cite{chavanis_12c}) it is argued, using such
approaches, that the characteristic time scale for collisional
relaxation has a generic scaling $\tau_{\rm coll} \sim N \tau_{dyn}$,
except for the special case of homogeneous QSS in one dimension.

In this paper we explore the conditions under which the
limit \eqref{coll_limit} is satisfied for the generic power law
interaction \eqref{pot}.
To do so we use a non-rigorous (but well 
defined) approach to the problem: we generalize the 
simple method initiated by Chandrasekhar
for the case of gravity \cite{chandra_42,binney}. 
This amounts to assuming that the dominant contribution 
to the collisionality, described by the  term $C_N$,  
comes from {\it two body collisions}.
For the gravitational interaction this simple approach
has turned out to account remarkably well for the 
observed time scales of collisional relaxation (in 
numerical simulations). We generalize
this approach to a generic power-law interaction; 
and compare the results obtained to the results
of numerical simulations of several such
systems. 

Several important results emerge
from this analysis. Firstly, it becomes evident
through this approach that, in general,
 the characteristic time $\tau_{\rm coll}$ for collisional relaxation 
scales with the particle number $N$ and may depend on the 
properties of the two body potential at small 
distances.   Our results for the two body collisional
relaxation lead to the conclusion that, in
this respect,  an important qualitative distinction
can be made between the cases 
$\gamma < d-1$ and $\gamma > d-1$:   
in both cases, for unsoftened potentials, $\tau_{\rm coll} \sim N^{\delta}$
where $\delta$ is a constant depending on
$\gamma$ and the dimension of space $d$.
However the sign of $\delta$ is positive
only if $\gamma < d-1$. This means 
that, when the size of the core is sent
to zero, the condition Eq.~(\ref{coll_limit})
can be satisfied only for $\gamma < d-1$.
The existence of QSS  requires the satisfaction
of this condition, and therefore such states
can exist for $\gamma \geq  d-1$ only if
the rate of collisionality is reduced through
the introduction of a  sufficiently soft core.  In other
words, for $\gamma < d-1$ QSS can
be considered to occur simply because
of the large distance behavior of the 
potential, while for  $\gamma \geq d-1$
their existence  depends on the details 
of the short-distance behavior.
This leads to what we call a  {\it dynamical}
(rather than thermodynamical) classification 
of the range of interactions, which has
been proposed also using different 
analyses in  
\cite{gabrielli_10, gabrielli_10b, chavanis_13, gabrielli_etal_2014}. 
 
The essential result above has already been reported in
\cite{gabrielli_10}. In this paper we present a 
more detailed and more extended study of collisional
relaxation in these systems, both for the analytical 
and numerical parts. In the analytical part we present
both a new quantitative treatment of the two body
relaxation including the contribution from hard collisions, 
and also of the case of different specified soft core
regularizations. In the numerical part we present 
much more extensive results and detailed analysis, 
including notably potentials which decrease more
slowly than the gravitational potential, and a
full quantitative exploration of the role of softening. The paper is organized 
as follows: in the next section we give a brief review on the
literature of the collisional relaxation in the context of
gravitational systems and detail our generalization of Chandrasekhar
calculation of the two body collisional relaxation rate for the pair
potentials \eqref{pot}, with soft or hard regularizations at small
distances.  This leads us to write parametric scalings
which allow us to infer our classification of the range of pair
interactions.  In the following section we describe the numerical
simulations we use to explore the validity of our analytical results,
their initial conditions and the macroscopic quantities we measure to
characterize collisional relaxation.  In the next section we present
our numerical results, first for the previously studied case of
gravity, and then for several cases with $\gamma>1$ and $\gamma<1$. We
compare then quantitatively the relaxation time obtained theoretically
with our simulations and, in the next section, we give numerical
evidence indicating that the maximum impact parameter scales with the
size of the system.  In the final section  we draw our conclusions.

\section{Relaxation rates due to two body collisions}
\label{sect-relaxation}
The parametric dependence of the characteristic time $\tau_{dyn}$ 
for mean field evolution is given by that of the 
typical time a particle needs to cross the system, of size $R$,  
under  the mean field force:  
\be
\label{tau_mf}
\tau_{dyn} \simeq \sqrt{\frac{m R^{\gamma + 2}}{g N}},
\ee
where $m$ is the mass of  each particle. The determination of the
parametric dependence of the characteristic time of collisionality $\tau_{coll}$  --- and, as expected, of relaxation towards thermodynamic 
equilibrium --- is much less evident. For the case of gravity ($\gamma=1$)
in three dimensions, Chandrasekhar gave the first estimates in 1943
\cite{chandrasekhar_43}, through a calculation of a diffusion coefficient 
in velocity space for an infinite homogeneous self-gravitating
distribution of particles. The central hypothesis, as for short-ranged systems, 
was to suppose that the main contribution to the collisional relaxation 
process arises from {\it two-body encounters}. He calculated the variation of velocity of a test particle undergoing a ``collision'' with 
a particle of the homogeneous distribution, the global relaxation process 
being the cumulative effect of such ``collisions''. As we will see in the 
next subsection the standard notion of {\it impact parameter} appears 
in the calculations. Due to the assumption of an infinite
homogeneous distribution and to the long-range nature
of gravity, Chandrasekhar had to cut-off the maximum impact 
parameter allowed at some scale, which he chose to be given
by the typical inter-particle separation.

More than twenty years after the paper of Chandrasekhar, H\'enon
\cite{henon_58} did a new calculation following the hypothesis of
Chandrasekhar, but considering that all the particles in the system
would contribute to the relaxation. There is then no need to introduce
artificially an upper cutoff in the impact parameter, as it is
naturally fixed by the size of the system. More recent theoretical
approaches, like e.g. \cite{weinberg_93, chavanis_10} (and references
therein), have followed a more complete approach, linearizing the
Boltzmann equation \eqref{vlasov}.  This approach makes possible to
take into account not only local but also collective effects. This
approach is, however, very cumbersome analytically and does not lead
in practice to definite conclusions about the issues we address here.

On the other hand, $N$-body computer simulations of the relaxation
problem have been performed to test the analytical predictions.  In
three dimensions such studies have been developed only for the case of
gravitational interaction.  We note, amongst others, numerical studies
focusing on the cosmological aspect \cite{diemand_04}, others focusing
on the maximum relevant impact parameter in the relaxation process
\cite{farouki_82,smith_92,farouki_94}. After some controversy, it
seems that the appropriate maximal impact parameter is the size
  of the system (rather than the inter-particle distance as postulated
  initially by Chandrasekhar). The study of the relaxation in
softened potentials (see e.g. \cite{theis_1998}) give more indications 
in this direction.  This is a result we will confirm and provide new evidence
for in this paper.

In the rest of this section we present our generalization of the two
body collisional relaxation time for any attractive power law
  pair potential of the form (\ref{pot}), with $\gamma >0$ and a soft
or hard core regularization at $r=0$, and any spatial dimension $d
\geq 2$.  The reasons for these restrictions on $\gamma$ and $d$
  become evident in the calculation below. These calculations give us
the parametric dependence for the relaxation rate via two body
collisions, $\Gamma=\tau_{\rm coll}^{-1}$, in a virialized system.  As
discussed in the introduction, if we assume that these processes are
the dominant ones in the collisional dynamics, we can then write the
condition for the existence of a regime in which a mean-field (Vlasov)
description of the dynamics is valid as \cite{prl_2010}
\be
\Gamma \, \tau_{dyn} \to 0 \quad {\rm when} \quad N \to \infty\,, 
\label{criterion-QSS}
\ee 
 Since QSS corresponds to the stationary (and thus virialized) states 
of the Vlasov equation, condition (\ref{criterion-QSS}) is also
a necessary one for the existence of such states. 
 
\subsection{{\bf Generalization of Rutherford scattering for generic power law interactions}}
\label{chandrasekhar}

We consider two particles of equal mass $m$, 
position vectors $\br_1$ and $\br_2$,
and velocity vectors  $\bv_1=\dot{\br}_1$ and 
$\bv_2=\dot{\br}_2$.
Their relative position vector is denoted
\be
\label{relative-vector}
\br=\br_1-\br_2
\ee
and their relative velocity $\bV=\dot \br$. 
In their center of mass frame, the velocities 
of the two particles are given by 
$\pm (\bV/2)$. Thus if $\Delta\bV$ is the 
change in the relative velocity of the 
particles in the two body encounter, 
the changes in velocity of the two 
particles in the laboratory frame,  
$\Delta \bv_1$ and $\Delta \bv_2$,  
(which are equal to those in the cent-re 
of mass frame) are
\bse
\begin{align}
\Delta\bv_1&=\frac{\Delta \bV}{2}\\
\Delta\bv_2&=-\frac{\Delta \bV}{2}.
\end{align}
\ese
The equations of the relative motion are those of
a single particle of mass $m/2$ with position
vector $\br(t)$ subject to the central potential.

\begin{figure}
  \begin{center}
          {\includegraphics[height= 0.2\textwidth]{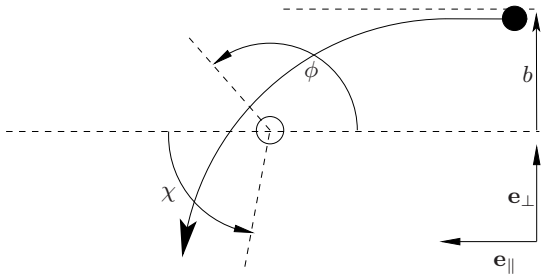}}
\caption{Trajectory of a particle in a two body collision in the 
cent-re of mass frame, with definition of the relevant quantities 
for its analysis, notably the deflection angle $\chi$.}  
\label{coll}
  \end{center}
\end{figure}

We decompose $\Delta\bV$ as 
\be
\Delta\bV=\Delta V_{\perp} {\mathbf e}_{\perp} + \Delta V_{\parallel} {\mathbf e}_{\parallel},
\ee
where  ${\mathbf e}_{\parallel}$ is a unit vector defined  parallel
to the initial axis of motion, and ${\mathbf e}_{\perp}$ a unit vector orthogonal
to it, in the plane of the motion (see Fig. \ref{coll}).
In the center of mass frame, the collision occurs as depicted in
Fig.~\ref{coll}, which shows the definition of the impact factor
$b$, and the deflection angle $\chi=2\phi-\pi$. 
As energy is conserved in the collision, the magnitudes 
of the initial and final relative velocity, $V=|\bV|$, are equal. 
It follows that 
\bse
\label{v_perp-and-v_par}
\begin{align}
\frac{\Delta V_\perp}{V} &= -\sin(\chi)\\
\frac{\Delta V_\parallel}{V} &= 1-\cos(\chi).
\end{align}
\ese
The angle $\phi$ can be calculated, as a function of the impact factor $b$, 
using the classic formula \cite{landau1}
\be
\label{phi-def}
\phi(b)=\int_{r_{min}}^{\infty} \frac{ (b/r^2) dr}{\sqrt{1 - (b/r)^2-4 v(r) / mV^2}},
\ee
where $r_{min}$ is  the positive root of the denominator. 

We consider now the case of a pure decaying
power law pair potential,  
\be
v(r)=-\frac{g}{r^\gamma}
\label{power-law-potential}
\ee
and $\gamma >0$. For $g>0$ the corresponding force is attractive,
while $g<0$ it is repulsive. In what follows we will consider 
the attractive case, but we will discuss below also the repulsive
case. Indeed it turns out that our essential results hold
in both cases.

The integral \eqref{phi-def} leads naturally to the definition of the
characteristic length scale
\be
\label{b0gamma}
b_0=\left(\frac{2|g|}{m V^2}\right)^{1/\gamma}\,.
\ee
Considering the attractive case, Eq.~\eqref{phi-def} may then
be rewritten as  
\be
\label{phi2}
\phi(b)=\int_{r_{min}}^{\infty} \frac{ (b/r^2) dr}{\sqrt{1 - (b/r)^2+2(b_0/r)^\gamma}}\,.
\ee
Changing to the variable $x=b/r$, we obtain
\be
\label{phi2-x}
\phi(b/b_0)=\int_0^{x_{max}} \frac{dx}{\sqrt{1 -x^2+2(b_0/b)^\gamma x^\gamma}},
\ee
 where now $x_{max}$ is the positive root of the denominator. 
Since $x_{max} $, for given $\ga$, is a function of $ b/ b_0$ 
only, it follows that $\phi$ is also a function of $b/ b_0 $ only. 

\begin{figure}
  \begin{center}
     {\includegraphics[height= 0.3\textwidth]{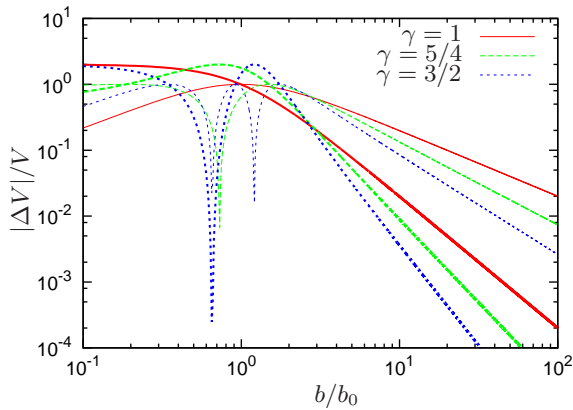}}
\caption{Absolute value of relative change 
in the perpendicular (thin lines) and parallel (thick
lines) components of the relative velocity in a two body 
encounter, for different attractive power-law potentials. 
The behaviors at small and large values of $b/b_0$ 
are well described by the analytical expressions 
given in the text.}
\label{ruth}
  \end{center}
\end{figure}

Equation \eqref{phi2} can be solved analytically only in a
few cases, and notably for the case $\gamma=1$ which corresponds to gravity in $d=3$.
For the general ($\gamma \neq 1$) case, the integral can easily
be computed numerically, and $\frac{\Delta V_\perp}{V}$ and 
$\frac{\Delta V_\parallel}{V}$ can then be calculated.   
Figure \ref{ruth} displays the results for a few chosen 
cases. In order to derive analytically the parametric dependences of 
the two body relaxation rate, it suffices, as we will see, to
have analytical approximations in the two asymptotic 
regimes of soft ($b/b_0 \gg 1$) and hard ($b/b_0 \ll 1$)
collisions. The corresponding expressions have been 
derived in a separate article \cite{chiron_15} by one
of us (BM) and another collaborator. In what follows
we make use of the relevant results of \cite{chiron_15},
where the full details of their derivations may be found.

\subsubsection{{\it Soft} collisions ($b \gg b_0$)}

When $b \gg b_0$ the particle trajectories are weakly
perturbed, and the collision is said to be {\it soft}.
It is shown in \cite{chiron_15} that, in this region,
one has 
\be
\label{douxmoins}
 \chi ( b / b_0 ) =  2A(\ga) (b_0/ b )^\ga 
 + \mathcal{O} ( (b_0/b)^{2 \ga} ) ,
\ee
where 
\be
\label{definition-A}
 A ( \ga ) = \sqrt{\pi} \frac{ \Gamma \left( \frac{ \ga +1 }{2} \right) }{\Gamma \left( \frac{\ga}{2} \right)}\,,
 \ee 
with $\Gamma(x)$ being the Euler Gamma function.
As the angle of deflection $\chi\ll 1$, it follows that 
\bse
\label{change-v-soft}
\begin{align}
\label{change-v-perp-soft}
\frac{\Delta V_\perp}{V} &=-2A(\gamma)\left(\frac{b_0}{b}\right)^\gamma +\mathcal{O}((b_0/b)^{2\gamma})\\
\label{change-v-par-soft}
\frac{\Delta V_\parallel}{V} &=2A(\gamma)^2\left(\frac{b_0}{b}\right)^{2\gamma}+\mathcal{O}((b_0/b)^{4\gamma}).
\end{align}
\ese
In Appendix \ref{alternative-derivation} an alternative 
derivation of Eq.~\eqref{change-v-perp-soft} is presented.

\subsubsection{{\it Hard} collisions ($b \ll b_0$)}
\label{sect-hard-coll}

It is shown in \cite{chiron_15} that, in this asymptotic regime, 
\be
\label{durplus1}
  \chi ( b / b_0 ) =  \, \frac{\gamma \pi}{2-\gamma}+\mathcal{O}\left((b/b_0)^\alpha\right),
\ee
where $\alpha = 2\gamma/(2-\gamma)$ for $\gamma<2/3$, $\alpha=b/b_0 \ln\left(b_0/b\right)$ for $\gamma=2/3$ and $\alpha=1$ for $2/3 <\gamma<2$. 
If $\ga \ge 2 $, collisions are well defined  with an
asymptotic free state \cite{chiron_15} only if  
\be
\label{condition-binary}
b > \beta b_0,
\ee
where
\be
\beta = \gamma^{1/\gamma}\left(1-\frac{2}{\gamma}\right)^{\frac{2-\gamma}{2\gamma}}.
\ee
For  $b \leq \beta b_0$, on the other hand, there is a finite time singularity, i.e.,
the relative distance of the particles vanishes at a finite time.

The first term in the asymptotic expansion Eq.~\eqref{durplus1} 
gives the angle of deflection in the limit of arbitrarily small
impact factors, and shows that it depends on $\gamma$.
While for the case $\gamma=1$ (i.e. gravity in $d=3$) 
each particle velocity is exactly reversed in the center 
of mass frame ($\chi=\pi$), the general result for the
deflection angle is different, and it increases to infinity
as $\gamma \rightarrow 2$ from below. At
$\gamma = 4/3$ each particle performs 
one full loop around the center of mass 
and escapes asymptotically in the same
direction it arrived in,  at $\gamma = 12/7$ 
each particle performs two full loops etc., 
and as $\gamma \rightarrow 2$ from below
the number of such loops diverges.

For $\gamma \ge 2$, as noted, there is in fact a singularity, with
the particles running into one another at a finite time. To include
this case in our treatment we must therefore assume that the pair 
potential Eq.~(\ref{power-law-potential}) is regularized at $r=0$,
so that there is a well defined collision for any impact factor.
It follows from our analysis that this means that the asymptotic
behavior below some arbitrarily small scale must be either repulsive,
or, if attractive, diverging more slowly that $1/r^2$. In what follows
this assumption will suffice to extend our results to the range
$\gamma \ge 2$.

\subsection{Computation of the cumulative effect of many collisions}
\label{many-coll}

Following Chandrasekhar we assume that thermal relaxation is induced by 
the randomization of particles velocity by two body collisions. 
In order to estimate the accumulated effect of two body collisions on
a particle as it crosses the whole system, we estimate first the number of 
encounters per unit of time with impact parameter $b$. In doing so
we make the following approximations:
\begin{enumerate}
\item the system is treated as a homogeneous random distribution of particles in a $d$ dimensional sphere of radius $R$, 
\item  the initial squared relative velocity of colliding particles is given by the variance of the particle velocities in the system.
\end{enumerate}

Each particle is then assumed to perform a simple homogeneous random walk in velocity space, with zero 
mean change in velocity (because the deflections due to each encounter have no preferred direction), and a positive mean squared velocity which we determine below. In this approximation, we assume that the particles have
rectilinear trajectories. This approximation clearly breaks down in the case of hard collisions, in which the trajectory is strongly perturbed. We expect however the estimation of the number of collisions per unit of time to remain
correct in this case, because encounters modify only the direction of the velocity, and not its modulus.
\begin{figure}
  \begin{center}
    {\includegraphics[height= 0.3\textwidth]{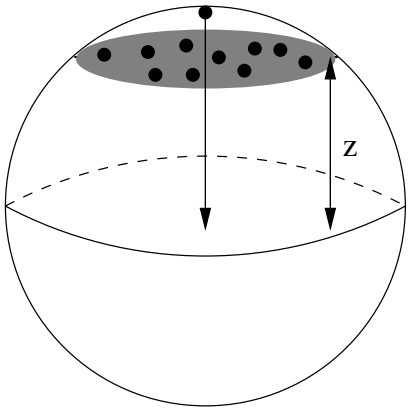}}
    \caption{The system is approximated as a perfectly spherical distribution 
of particles  with radius $R$.} 
\label{derivation}
  \end{center}
\end{figure}

As illustrated schematically in Fig.~\eqref{derivation}, we now 
divide the system in disks of thickness $dz$, and write the
average number of encounters with impact parameter 
between  $b$ and $b+db$ of a particle crossing this 
disk as 
\be
\delta n=\frac{B_d N}{R^d} b^{d-2}\, db\, dz
\label{eq22}
\ee
where $B_d$ is a numerical factor which depends on the 
spatial  dimension $d$ (e.g. $B_2=2/\pi$, $B_3=3/2$). 

Multiplying Eq.~\eqref{eq22} by the square of Eq.~\eqref{v_perp-and-v_par} with the condition \eqref{phi2-x}, and integrating from $z=0$ to $z=R$ and from
$b=0$ to $b=\sqrt{R^2-z^2}$, we then estimate the average change in the  
velocity {\it during one crossing} of the system, for the perpendicular and  
parallel components of the velocity respectively, as:
\be
\label{total-change-v}
\frac{\langle|\Delta V_{\perp,\parallel}^2|\rangle}{|V^2|} = 2B_d N \left(\frac{b_0}{R}\right)^{d-1} 
\,{\mathcal I}_{\perp, \parallel} \left(\frac{b_0}{R}\right)
\ee
where 
\be
\label{I-powerlaw}
{\mathcal I}_{\perp, \parallel} (x_R)= \int_0^{x_R} dx \,x^{d-2} \, \Theta_{\perp, \parallel} (x) \,\sqrt{1-\frac{x^2}{x_R^2}}
\ee
where $x=b/b_0$, $x_R=R/b_0$ and
\bse
\label{theta}
\begin{align}
\Theta_\perp(x)&=\sin^2\left(\chi(x)\right)\\
\Theta_\parallel(x)&=\left[1-\cos(\chi(x))\right]^2.
\end{align}
\ese
Writing the expression for $\frac{\langle|\Delta V_{\perp,\parallel}^2|\rangle}{|V^2|}$ 
in this way allows a simple and useful comparison
with the case of particles interacting by an exact repulsive 
hard core potential. Indeed it is straightforward to show 
(see e.g. \cite{balescu_97}) that for (infinitely) hard 
particles with a diameter $\sigma$, one has 
\be
\label{deflection-hardcore}
\chi (b) = \left\{ 
\begin{array}{ll}
2\arccos\left(\frac{b}{\sigma}\right)   & \text{if } b \leq \sigma \\
 0 & \text{otherwise.}
\end{array}
\right.
\ee
Calculating $\frac{\langle|\Delta V_{\perp,\parallel}^2|\rangle}{|V^2|}$
for this case using exactly the same approach used above, one
obtains, for the case $\sigma=b_0$, exactly Eq.~\eqref{total-change-v} 
with 
\bse
\begin{align}
{\mathcal I}_{\perp}&=\frac{8}{(d+3)(d+1)}\\
{\mathcal I}_{\parallel} &=\frac{4}{d-1} {\mathcal I}_{\perp}.
\end{align}
\ese

Let us return now to the expressions  Eq.~\eqref{I-powerlaw}
for the case of (attractive) power law interactions.  Given that 
$x_R \gg 1$ we can make the approximation
\bea
\label{int-divided}
 {\mathcal I}_{\perp, \parallel} (x_R) &&\approx \int_0^{1} dx \,x^{d-2} \, \Theta_{\perp, \parallel} (x) \nonumber \\
 &+&\int_1^{x_R} dx \,x^{d-2} \, \Theta_{\perp, \parallel} (x) \,\sqrt{1-\frac{x^2}{x_R^2}}.
 \eea
 
The first integral gives the contribution due to hard
collisions ($b <b_0$). It is finite provided only  
that the deflection angle is well defined, i.e., 
provided only that the two body collisions is well 
defined. As we have discussed above this is
true for any $\gamma < 2$, and for $\gamma  \geq 2$ 
if we assume the singularity at $r=0$  to be
appropriately regularized.  Thus this term gives
a contribution to $\frac{\langle|\Delta V_{\perp,\parallel}^2|\rangle}{|V^2|}$
which has precisely the parametric dependences of 
an exact repulsive hard core, differing only by an
overall numerical factor. 

 Considering now the second term, giving the contribution
 from soft collisions ($b >b_0$),  we see that there are 
 two different cases according to the large $x$ behavior of 
 $\Theta_{\perp,\parallel}$: the integral is convergent 
 as $x_R \rightarrow \infty$ if and only if 
 $x^{d-1} \Theta_{\perp,\parallel} (x) \rightarrow 0$ as 
 $x \rightarrow \infty$. We thus infer from Eqs.~(\ref{change-v-soft})
 the following: 
  
 \begin{itemize}

\item For  $0 < \gamma < (d-1)/2$, 
\bea
\label{I-vlr}
 {\mathcal I}_{\perp} (x_R)   &\approx& 4A^2(\gamma)  \int_0^{x_R} dx \,x^{d-2-2\gamma}  \,\sqrt{1-\frac{x^2}{x_R^2}} \\
                                               &=& A^2(\gamma) \sqrt{\pi} \frac{\Gamma\left[d/2-1/2-\gamma\right]}{\Gamma\left[d/2+1\gamma\right]}  x_R^{d-1-2\gamma}
\eea
 and  ${\mathcal I}_{\parallel} (x_R) \ll  {\mathcal I}_{\perp} (x_R)$. Thus the integral
 is dominated by the contribution of soft scatterings, for which the change
 in the relative velocity is predominantly orthogonal to the initial relative velocity.  
Replacing Eq.~\eqref{I-vlr} in Eq.~\eqref{total-change-v}, we obtain the scaling
\be
\label{scaling-slr}
\frac{\langle |\Delta \bV^2| \rangle}{V^2} \approx \frac{\langle|\Delta V_{\perp}^2|\rangle}{|V^2|}\sim N\left(\frac{b_0}{R}\right)^{2\gamma}.
\ee
where 
\be
\frac{\langle|\Delta \bV|^2\rangle}{V^2}=\frac{\langle|\Delta \bV_{\perp}|^2\rangle}{V^2}+\frac{\langle|\Delta\bV_{\parallel}|^2\rangle}{V^2}.
\ee

\item For $\gamma = (d-1)/2$, which corresponds to gravity
in $d=3$, the contribution from all impact factors from 
the scale $b_0$ must be included and 
\be
 {\mathcal I}_{\perp} (x_R)   \approx 4A^2(\gamma)  \ln \, x_R\,.
 \ee
As in the previous case, ${\mathcal I}_{\parallel} (x_R) \ll  {\mathcal I}_{\perp} (x_R)$.
Note that, given $x_R \gg 1$ this result for ${\mathcal I}_{\perp} (x_R)$ is very insensitive 
to precisely where the lower cut-off at $b \sim b_0$ is chosen.  
We obtain therefore
\be
\label{scaling-grav-like}
\frac{\langle |\Delta \bV^2| \rangle}{|V^2|} \sim N\left(\frac{b_0}{R}\right)^{d-1}\ln\left(\frac{R}{b_0}\right).
 \ee

\item For  $\gamma>(d-1)/2$,  we have 
\be
 {\mathcal I}_{\perp, \parallel} (x_R) \approx   {\mathcal I}_{\perp, \parallel} (\infty) \approx \int_0^{\infty} dx \,x^{d-2} \, \Theta_{\perp, \parallel} (x)
 \ee
which is a constant that can be numerically calculated in a straightforward way for any given pair 
potential in this class. We obtain therefore
\be
 \label{scaling-mlr}
 \frac{\langle|\Delta \bV^2|\rangle}{V^2}\sim N\left(\frac{b_0}{R}\right)^{d-1}.
 \ee

\end{itemize}

In the last case, for sufficiently rapidly decaying potentials, we obtain therefore 
the same scaling as for the case of hard core particles of diameter $b_0$.  

\subsection{Scalings with $N$ of the relaxation rate in a QSS}

\label{rel_rate}
Using these results, we now determine how the relaxation rate scales 
with the parameters of the system. Assuming the system to be in a 
QSS we can then obtain its scaling as a function of $N$ alone.  
For clarity we drop irrelevant numerical prefactors, but these will be
analyzed further in Sect.~\ref{quantitative}. 

We define the relaxation rate $\Gamma$ as the inverse of the time
scale at which the normalized average change in velocity squared due to 
collisions is equal to one. Given that the estimated 
$\frac{\langle|\Delta \bV|^2\rangle}{V^2}$ is the average 
change in a crossing time $\tau_{dyn}$, we have therefore
\be
\label{def_gamma}
\Gamma\tau_{dyn} \simeq \frac{\langle|\Delta \bV|^2\rangle}{V^2}\,.
\ee

In order to obtain the scaling with $N$ from the above results, we need to
determine how the ratio $b_0/R$ scales with $N$. Using the definition
Eq.~\eqref{b0gamma} and assuming, as stated above, that  
the modulus of the relative velocity of colliding particles 
can be taken to be of the same order as the typical 
velocity of a single particle $v$, we have 
\be
\left(\frac{b_0}{R}\right)^\gamma\sim\frac{g}{mv^2 R^\gamma} \sim \frac{1}{N} \frac{gN^2}{(mNv^2) R^\gamma} 
\sim \frac{1}{N} \frac{U}{K} 
\label{scaling-virialization}
\ee
where $K$ is the total kinetic energy and $U$ the total potential energy of the system.

If we now assume the system to be in a QSS, i.e. in virial equilibrium, 
the {\it virial theorem} gives that
\be
\label{virial}
2K +\gamma U = 3 PV,
\ee
where $P$ is the pressure of the particles on the boundaries if the system is enclosed, 
and $P=0$ if the system is open.

By definition the mean-field scaling with $N$ makes each term in Eq.~\eqref{virial}
scale in the same way with $N$ so that the relation remains valid independently
of $N$ (up to finite $N$ fluctuations). Thus using this scaling we can infer that
\be
\label{scaling-bmin}
b_{0}\sim R N^{-1/\gamma}.
\ee

Using Eqs.~\eqref{scaling-slr}, \eqref{scaling-grav-like} and \eqref{scaling-mlr}, we
then infer the following behaviors:
\begin{itemize}
\item For $0<\gamma<(d-1)/2$, 
\be
\label{rate_llr}
\Gamma \,\tau_{dyn}  \sim N (b_0/R)^{2\gamma}\sim N^{-1}.
\ee 
\item For $\gamma=(d-1)/2$
\be
\label{rate_gr}
\Gamma\,\tau_{dyn} \sim N^{-1} \ln \left(N\right).
\ee

\item For $\gamma>(d-1)/2$   
\be
\label{rate_slr}
\Gamma \,\tau_{dyn}\sim N^{-(d-1-\gamma)/\gamma}.
\ee

\end{itemize}

It follows that that the condition Eq.~\eqref{criterion-QSS} only holds for potentials with $\gamma < d-1$.
Only in this case therefore can the QSS be supposed to 
exist as we have assumed.  For $\gamma \geq d-1$, on the other 
hand, the relaxation induced by two body collisionality 
occurs on a time scale which is short compared to
a particle crossing time, and a stationary non-thermal
state cannot exist on the latter time scale, i.e., a 
QSS cannot exist.

\subsection{Relaxation rates for softened power-law potentials}
\label{theo-soft}

We consider now the case in which the power-law potential is
``softened" at short distances, i.e., regulated so that the modulus 
of the force between two particles is bounded above at some 
finite value.  The principle motivation for considering this case 
here is that, in practice, even for $\gamma <2$, we are unable 
numerically to test directly the validity of the scaling 
predictions Eqs.~\eqref{scaling-mlr}-\eqref{rate_llr} 
for the exact (singular) potentials: 
the numerical cost of integrating sufficiently accurately 
hard two body scatterings over the long time scales 
required is prohibitive. Instead we will consider power-law
potentials softened at a scale $\epsilon$, and 
study the scaling with both $N$ and $\epsilon$
of the relaxation rates in the numerically accessible
range for these parameters. 

A detailed analysis  of the two body scattering for such softened power law 
potentials has been given also in \cite{chiron_15}. We again use the results of this
paper to infer, using Eqs.~\eqref{total-change-v}-\eqref{theta} above, the 
parametric scalings of the relaxation rate. As in the previous section, we 
defer until later a discussion of the exact numerical factors, for the 
specific smoothing functions used in our numerical simulations.   

As softening modifies the force below a characteristic scale $\epsilon$, its 
effect is to modify the deflection angles for impact factor $b$ below a 
scale of the same order. From the considerations above it is then
evident that, for $\epsilon < b_0$, such a softening does not change
the parametric scalings: it can only change the numerical value 
of the (finite) first integral in Eq.~\eqref{int-divided}. 
For $\epsilon > b_0$, on the other hand, the second integral 
in Eq.~\eqref{int-divided} is modified because the functions
$\Theta_{\perp,\parallel}$ are modified up 
to $x \sim \epsilon/b_0$. Assuming that $\epsilon \ll R$,
this will lead to a modification of the parametric scaling of the 
full expressions for  $\frac{\langle|\Delta \bV|^2\rangle}{V^2}$
when $\gamma \geq (d-1)/2$.  In \cite{chiron_15}  it is shown
that, when $\epsilon \geq b_0$, the deflection angle can be 
approximated as
\be
\label{phi-approx-large-eps}
\chi \simeq \left\{ 
\begin{array}{ll}
2B(\ga) \left(\frac{b_0}{\ep}\right)^{\gamma}\left(\frac{b}{\ep}\right) & \text{if } b < \ep^* \\
2A(\ga) \left(\frac{b_0}{b}\right)^\ga & \text{if } b > \ep^*,
\end{array}
\right.
\ee
where $B(\gamma)$ is a finite constant  the exact value of which depends 
on the functional form of softening used (and $A(\ga)$ is as defined in
Eq.~\eqref{definition-A}). The scale $\ep^*$ is of the same order 
as $\epsilon$ (from continuity of Eq.~\eqref{phi-approx-large-eps} at 
$b=\ep^*$, their ratio is given by $\epsilon^*/\epsilon\sim(A/B)^{\frac{1}{1+\ga}}$).

Using Eq.\eqref{phi-approx-large-eps} we can now calculate approximately
the second integral in  Eq.~\eqref{int-divided} for the cases in which the 
parametric dependence of their values are modified by the smoothing
(with $\epsilon > b_0$):

\begin{itemize}
\item For $\gamma > (d-1)/2$ (taking $x_R\to\infty$):
\bse
\begin{align}
{\mathcal I}_{\perp}&\simeq \left[\frac{B^2(\gamma)}{d+1}+\frac{A^2(\ga)}{2\ga-d+1}\right]\left(\frac{\ep}{b_0}\right)^{d-1-2\ga} \label{I-smoothed-perp} \\
{\mathcal I}_{\parallel}&\simeq \left[\frac{B^4(\gamma)}{4(d+3)}+\frac{A^4(\ga)}{4\ga+1-d}\right]
\left(\frac{\ep}{b_0}\right)^{d-1-4\ga} \label{I-smoothed-par}.
\end{align}
\ese
and therefore ${\mathcal I}_{\perp} \gg {\mathcal I}_{\parallel}$ if $\epsilon \gg b_0$.

\item For $\gamma=(d-1)/2$, assuming $x_R \gg (\epsilon/b_0)$ (i.e. $\epsilon \ll R$), 
we obtain
\be
{\mathcal I}_{\perp}\simeq A^2(\gamma) \ln\left(\frac{R}{\ep}\right)\,,
\ee
while ${\mathcal I}_{\parallel}$ is given as Eq.~\eqref{I-smoothed-par}, and
${\mathcal I}_{\perp} \gg {\mathcal I}_{\parallel}$ if $\epsilon \gg b_0$.
\end{itemize}

Using these results we infer finally that the scalings of the relaxation rates 
of a QSS (with $b_0$ scaling as in Eq.~ \eqref{scaling-bmin})  in the 
large $N$ limit are the following:
\begin{itemize}
\item If $0<\gamma<(d-1)/2)$, 
\be
\label{rate_llr-soft}
\Gamma \,\tau_{dyn}  \sim N^{-1},
\ee 
i.e. the same as in the absence of smoothing; 
\item If $\gamma>(d-1)/2$, then   
\be
\label{rate_slr-soft}
\Gamma \,\tau_{dyn}\sim N^{-1} \left(\frac{\epsilon}{R}\right)^{d-1-2\gamma}  
\ee

\item If $\gamma=(d-1)/2$, then
\be
\label{rate_gr-soft}
\Gamma \,\tau_{dyn} \sim N^{-1}\ln\left(\frac{R}{\epsilon}\right).
\ee

\end{itemize}

In summary, the correct parametric scaling for 
the two body relaxation rates of a QSS, in the case
of a power-law potential softened at a scale 
$\epsilon > b_0$, are well approximated by 
simply introducing a cutoff at an impact 
factor of order $\epsilon$ (and therefore
considering only the contribution from 
soft collisions). 

For what concerns the existence of QSS, we thus
conclude that, with a softened power law potential, 
one can satisfy  the condition Eq.~\eqref{criterion-QSS} 
even for any $\ga \geq d-1$. Indeed, taking $\epsilon/R$
to be independent  of $N$ (i.e. scaling the softening
with the system size), we obtain in all cases that
$\Gamma_\epsilon\,\tau_{dyn} \sim N^{-1}$. More generally, it is straightforward to deduce 
what scaling of $\epsilon$ with $N$ is required
to satisfy the condition Eq.~\eqref{criterion-QSS}
in the mean-field limit.

\section{Numerical simulations}

We have performed numerical simulations in $d=3$ of the evolution of $N$ particle
systems, extending to sufficiently long times to observe their collisional 
evolution\footnote{For a recent numerical study of these systems focusing on the 
shorter time (mean field) evolution i.e. collisionless relaxation, see \cite{DiCintio_etal_2013, DiCintio_etal_2015}.}. As we have discussed in the previous section, exact power law 
interactions with $\gamma \ge (d-1)/2$ lead to strong collisions at impact factors $b<b_0$.
Indeed, as we have seen, when $\gamma$ increases much above unity
particles can even make multiple loops around one another during 
collisions (cf. Eq.~\eqref{durplus1}). The smaller is $b$, the shorter is the 
characteristic time for a collision compared to the mean field
time and therefore the greater is temporal resolution required for an accurate
integration (and, in particular, conservation of the energy). This means
it is too expensive numerically, even for a few thousand
particles, to accurately simulate such a system for times long 
enough to be comparable to the predicted relaxation times. Indeed 
we have seen that the calculation we have done predicts that,
even for $(d-1)/2 < \gamma < d-1$ (i.e. $1<\ga<2$ in $d=3$), relaxation
should be dominated by strong collisions with $b \sim b_0$ but
nevertheless $\Gamma \,\tau_{dyn}$ diverges in the mean
field limit.

For these reasons, we employ a potential with a softening which
is sufficiently large to suppress strong collisions. The predicted scalings 
we can test are thus those given  in Sect.~\ref{theo-soft}, rather than 
the ones corresponding to pure power-law potentials given in 
Sect.~\ref{rel_rate}. By studying also the scalings with the softening
$\epsilon$ at fixed $N$, however, we can indirectly test in this
way the extrapolation to the scalings in Sect.~\ref{rel_rate}. 

\subsection{Code}

We use a modification of the publicly available gravity code GADGET2\cite{gadget}. 
The force is computed using a modified Barnes and Hut tree algorithm, and we 
have modified the code in order to treat pair potentials of the form
Eq.~\eqref{pot} and softened versions of them (which are those
we use in practice). We use an opening angle $\theta=0.001$, which
ensures a very accurate computation of the force. The evolution of
  the system is computed using a Verlet-type Drift-Kick-Drift
  symplectic integration scheme. The simulations are checked using
  simple convergence tests on the numerical parameters, and their
  accuracy is monitored using energy conservation. For the time-steps
  used here it is typically conserved to within $0.1\%$ over the whole run,
  orders of magnitude smaller than the typical variation of the
  kinetic or potential energy over the same time.

\subsection{Initial and boundary conditions}

As initial conditions we take the $N$ particles randomly distributed 
in a sphere of radius $R=1/2$, and ascribe velocities to particles so that
each component is an independent uniformly distributed variable in an
interval $[-\xi,\xi]$ (i.e.  ``waterbag" type initial conditions in
phase space). The parameter $\xi$ is chosen so that initial virial
ratio is unity, i.e., $2K/|U|=\gamma$. We make this choice of initial
conditions because it is expected to be close to a QSS,  to
which (collisionless) relaxation should occur ``gently", and this
is indeed what we observe. We have chosen to enclose the system 
in a cubic box of size $L=1$, in order to avoid the complexities
associated with particle evaporation. This constraint  is
imposed in practice using {\it soft} boundary conditions, which are
implemented by changing the sign of the $i^{th}$ component of the
velocity when the $i^{th}$ component of the position lies
outside the simulation box. We use a time step of the order of
$10^{-3}\tau_{dyn}$ (which provides well converged results),
where $\tau_{dyn}$ is defined precisely below.

\subsection{Softening}
\label{softenings}
We have performed simulations using two different softening
schemes: a ``compact'' softening and a ``Plummer'' softening.
The former corresponds to a two body potential  
\be
\label{compact-potential}
v^{\text{C}}(r,\epsilon) = \left\{ 
\begin{array}{ll}
-\frac{g}{r^\gamma } & \text{if } r \geq \ep \\
- \frac{g}{\ep^\ga} \tv \left (r/\ep \right) & \text{if } 0 \leq r \leq \ep,
\end{array}
\right.
\ee
where  $\tv (x)$ is a polynomial, of which the exact expression is given in 
App.~\ref{appendix_potential}. It is chosen so that the potential and its first two derivatives are continuous at $r=\epsilon$, and it interpolates
to a force which vanishes at $r=0$ via a region in which the force
becomes repulsive.
The Plummer smoothing corresponds to the simple potential
\be
 v_\epsilon^{P}(r)=-\frac{g}{(r^2+\ep^2)^{\gamma/2}},
 \ee
which is everywhere attractive. 

\begin{figure}
\begin{center}
    {\includegraphics[height= 0.35\textwidth]{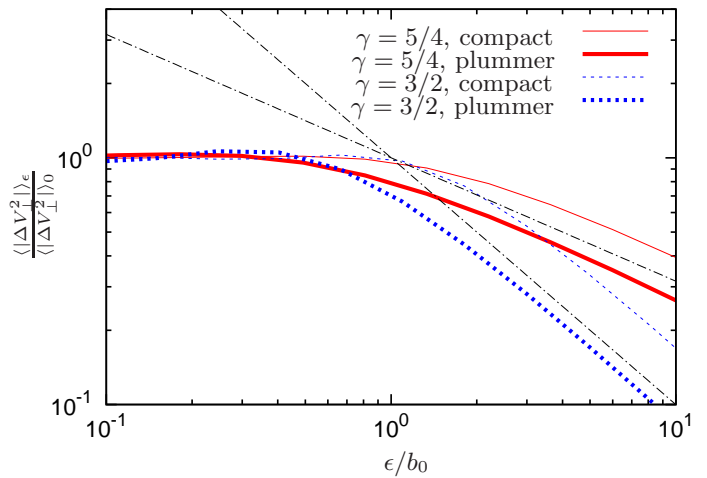}}
\caption{Numerical evaluation of Eqs.~\eqref{phi-def} and Eq.~\eqref{total-change-v} normalized to the value for $\epsilon/b_0\to0$ for $\gamma=5/4$ and $\gamma=3/2$. The power-law lines are the theoretical scaling \eqref{rate_slr-soft}. } 
\label{rutherford_softening3}
  \end{center}
\end{figure}

As we have noted it is straightforward to calculate numerically
the relaxation rates for these softened potentials, using 
Eq.~\eqref{phi-def} and Eq.~\eqref{total-change-v}.
We show in Fig.~\ref{rutherford_softening3} the ratio
of the resultant $\frac{\langle| \Delta \bV_{\perp}^2|\rangle}{\langle|V^2|\rangle}$ 
compared to its value for the exact power law, for $\gamma = 5/4$ and $\gamma=3/2$,
as a function of the ratio $\epsilon/b_0$. As described in the
previous section we observe that, for $\epsilon\ll b_0$, the effect 
of the softening is negligible, while for $\epsilon \gg b_0$,  we 
recover a simple power law scaling with $\epsilon$ which 
agrees with that derived above for this regime, cf. Eq.~\eqref{rate_slr-soft}.
We note that in Fig.~\ref{rutherford_softening3} the 
normalization for the asymptotic Plummer curves is greater 
that for the compact softening. 

Performing simulations with these two different softening
schemes allows us to test not just the robustness of the agreement
with the theoretical scalings derived above, which should not depend
on the details of the softening scheme. It also allows to test more
quantitatively for the correctness of the theoretical predictions for
the relaxation rates, which predicts also the relative amplitude of
the relaxation rate in the regime $\epsilon \gg b_0$. To facilitate
this comparison it is convenient to define an effective softening
$\ep_{eff}$ obtained by assuming that all the collisions are soft,
i.e.,
\be
\label{phi-approx-large-eps-estimate}
\chi_\ep \simeq \left\{ 
\begin{array}{ll}
0 & \text{if } b < \ep_{eff} \\
 2A(\ga)\left(\frac{b_0}{b}\right)^\ga & \text{if } b \ge \ep_{eff}.
\end{array}
\right.
\ee
Computing the same quantity as in Fig.~\eqref{rutherford_softening3}, we
can determine, by matching with the result for any other softening scheme,  
a value of $\ep_{eff}$ in units of $\epsilon$. We can compute therefore an effective softening using
\be
\label{alpha-soft}
\ep_{eff}=\alpha \,\ep \,,
\ee
where the values of $\alpha$ are given in Tab.~\ref{table-soft} for
  our two softening schemes, for the values of $\gamma$ we explore
  here (in the range $\gamma \geq 1$ where the softening plays a
  role).  The result for the case of gravity and the Plummer softening
  is in agreement with that derived in \cite{theis_1998} (see also
  \cite{white_78}).

Thus our analytical calculations predict that the relaxation rates
of QSS measured with the different softening schemes should
not only scale in the same way as a function of $\epsilon$
(for $\epsilon \gg b_0$) but also they should be equal at
values of $\epsilon$ corresponding to the same $\epsilon_{eff}$. 

\begin{table}
\begin{tabular}{|c|c|c|}
\hline
$\gamma$ & compact core & plummer core \\
\hline
1   & 0.80  & 1.69\\
5/4 & 0.74  & 1.55 \\
3/2 & 0.75  & 1.50\\

\hline  
\end{tabular}
\caption{Factor $\alpha$ (see Eq.~\eqref{alpha-soft}) to compute the effective softening $\epsilon_{eff}$ (see text) in units of $\epsilon$, for the two different
softening schemed used in this work.}
\label{table-soft}
\end{table}
\subsection{Sets of simulations}

We performed, for each value of $\gamma$, and each softening scheme,
two different kinds of sets of simulations. One set is at fixed particle number 
$N$ and a range of different values of the softening $\epsilon$, 
while in the other set $\epsilon$ is kept constant and $N$ is varied.  
To refer to the simulations we will use the notation 
$\mathcal C(\gamma;N,\epsilon)$ for a simulation 
with the compact (``$\mathcal C$'') softening \eqref{pot}, 
power law exponent $\gamma$, particle number $N$ and softening
$\epsilon$. Similarly we denote  $\mathcal P(\gamma;N,\epsilon)$
a set of simulations with the Plummer (``$\mathcal P$'') smoothing.

The simulations on which our results below are based are the
following:
\begin{itemize}

\item A set $\mathcal C(\gamma ;N = 8000,\epsilon)$ for $\gamma = 1/2$, $\gamma = 1$ , $\gamma = 5/4$ and $\gamma =3/2$  with the values of $\epsilon$ listed in the first column of Tab.~\ref{table-sim}.

\item A set $\mathcal C(\gamma ;N, \epsilon/L = 0.005)$ for $\gamma = 1/2$, $\gamma = 1$ , $\gamma = 5/4$ and $\gamma =3/2$  with the values of $N$ listed in the third column of Tab.~\ref{table-sim}. 

\item A set $\mathcal P(\gamma ;N = 8000,\epsilon)$ for  $\gamma = 5/4$ and $\gamma =3/2$  with the values of $\epsilon$ listed in the second column of Tab.~\ref{table-sim}.

\item A set $\mathcal P(\gamma ;N, \epsilon/L = 0.005)$ for  $\gamma = 5/4$ and $\gamma =3/2$  with the values of $N$ listed in the third column of table Tab.~\ref{table-sim}.

\end{itemize}

\begin{table}
\begin{tabular}{|c|l|}
\hline
$\epsilon/L$ &  0.0005\, 0.001 \, 0.002 \, 0.003 \, 0.004 \, 0.005\\
& 0.01 \, 0.02 \, 0.03 \, 0.04  \\
\hline
$N$  & $10^3$ \, $12^3$ \, $16^3$ \, $20^3$ \, $26^3$ \, $30^3$ \\
 \hline  
\end{tabular}
\caption{List of simulations: The first row gives the values of
the softening parameter $\epsilon$ used in two sets of simulations
with $N=8000$ particles; the second column gives the values of $N$
employed in two sets of simulations at fixed $\epsilon/L=0.005$.}
\label{table-sim}
\end{table}

\subsection{Numerical estimation of the relaxation rate}
To measure numerically the relaxation rate of a QSS we study
the temporal evolution of different quantities. We consider
principally two quite different quantities. On the one hand
the total kinetic (or potential) energy of the system, 
and on the other hand, the averaged quantity 
defined as
\be
\label{estimate_micro}
\Delta(t)\equiv\frac{\langle (e(t)-e(t^*))^2\rangle}{2k^2(t^*)}\,, 
\ee
where $e(t)$ is the total energy of a single particle (at time $t$), 
and $k(t)$ is the kinetic energy per particle. The time $t^*$
is an initial chosen time (and thus $t > t^*$) at which the
system has relaxed, starting from the initial condition, to
a QSS (typically we have below of $t^* \sim 10\,\tau_{dyn}$).
The brackets $\left<\cdot\right>$ indicate an average 
over all the particles in the system.

While the first quantity probes simply the macroscopic
evolution of the system in a ``blind" manner, the second 
quantity probes more directly the microscopic evolution 
of the quantities considered in the theoretical calculation.
Indeed the calculation in Sect.~\ref{chandrasekhar} 
provides a prediction for the average variation of 
the velocity of particles due to collisions. The difficulty
with measuring this directly is that the velocity of
particles also changes also continuously because of the
mean field potential. Particle energy, on the other hand,
remains exactly constant in a QSS, and its change
is in principle due to collisional effects, which we posit 
here are dominated by the two body collisions. 

\subsection{Other indicators of relaxation}

In order to determine whether the system is in a QSS
state (and hence not in thermal equilibrium), we compute 
moments of the system's velocity distribution. If the system is at
thermal equilibrium, the probability distribution of velocities  
must be Gaussian for each component with zero mean,  and therefore  
all odd moments of such components must vanish, while even moments of 
order higher than two are determined as a simple power of the variance: 
\[
\langle v_i^{2n} \rangle = (2n-1)!!\langle v_i^2 \rangle^{n}\,,
\]
In order to detect the deviation from Gaussianity of the velocity distribution 
we use the first two even moments of order larger than two, normalized
so that they are zero in the case of Gaussianity:
\bea
\phi_4&=&\frac{\overline{\langle v_i^4 \rangle}}{3\overline{\langle v_i^2 \rangle}^2}-1\\
\phi_6&=&\frac{\overline{\langle v_i^6 \rangle}}{15\overline{\langle v_i^2 \rangle}^3}-1\,,
\eea
where $\overline{\cdot}$ denotes average over the coordinates.

\subsection{Units}

As noted above we take the side of the enclosing box $L=1$. 
The mean field characteristic time is defined (following Eq.~\eqref{tau_mf})
as:
\be
\label{tau-mf}
\tau_{dyn}=\sqrt{\frac{m L^{\gamma+2}}{g N}}\,
\ee
and we report our results for velocities in units of 
\be
\label{vel-norm}
v^*=\frac{L}{\tau_{dyn}}=\sqrt{\frac{g N}{m L^{\gamma}}}.
\ee

\section{Results for case of gravity ($\gamma=1$)} 
\label{simu-grav}
\begin{figure*}
\begin{center}
\subfloat[]{\label{N1_pot_eps_some}{\includegraphics*[width=8cm]{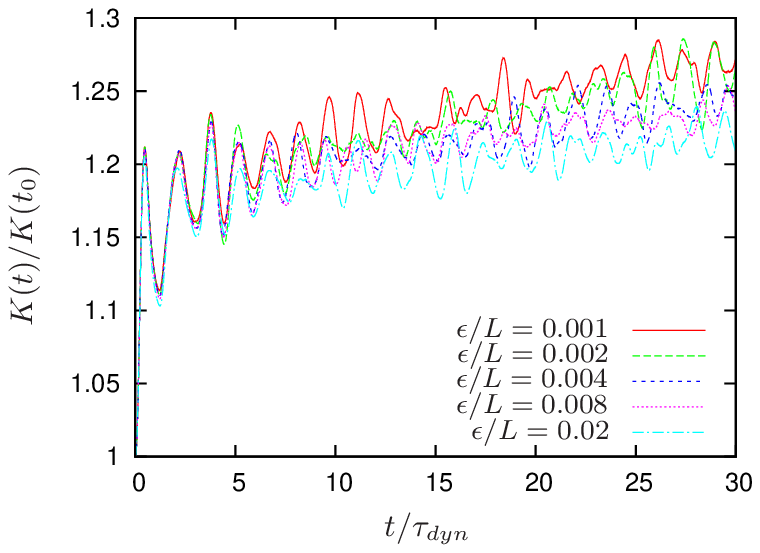}}}
\subfloat[]{\label{N1_pot_N}{\includegraphics*[width=8cm]{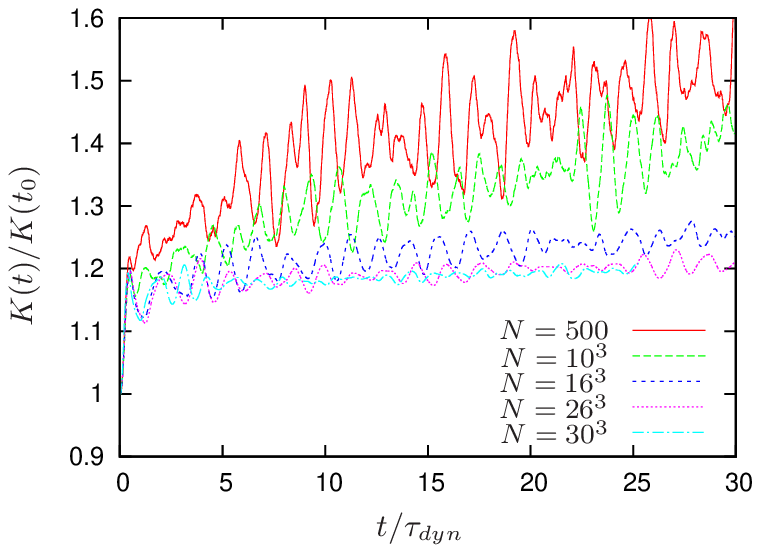}}}\\
\subfloat[]{\label{N1_L1_vel_425_ep0.001}{\includegraphics*[width=8cm]{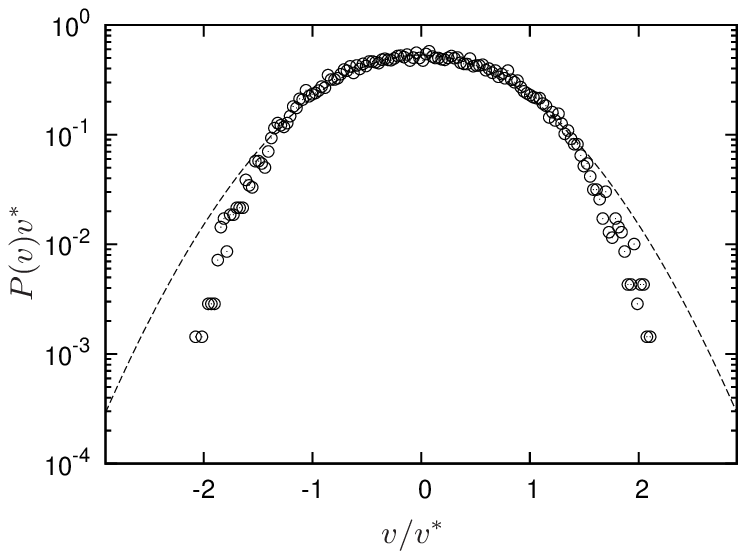}}}
\subfloat[]{\label{phi_N1_L1}{\includegraphics*[width=8cm]{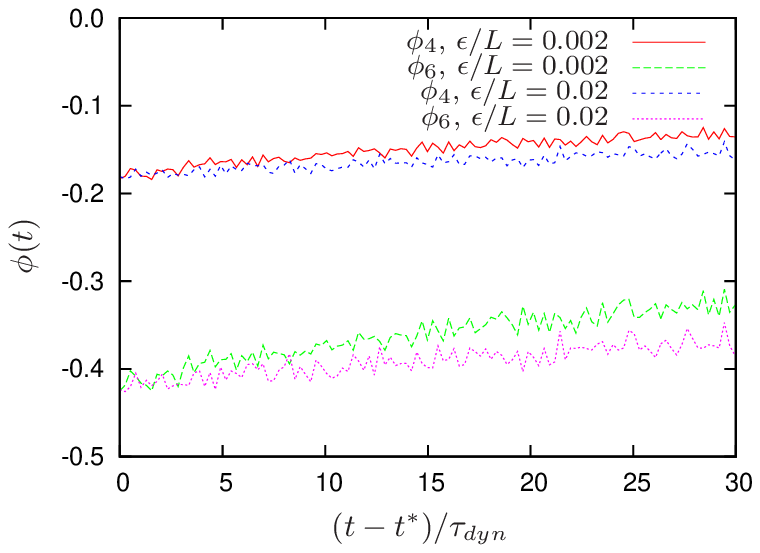}}}\\
\subfloat[]{\label{N1_rho_eps}\includegraphics*[width=8cm]{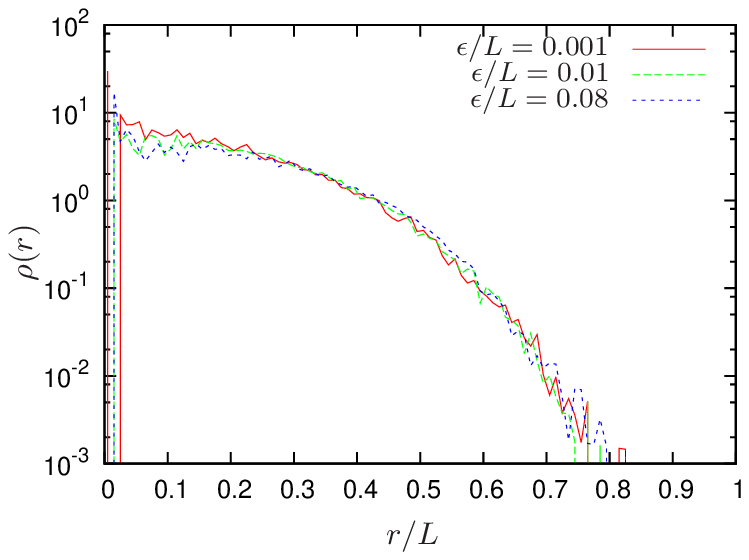}}
\subfloat[]{\label{N1_rho_N}{\includegraphics*[width=8cm]{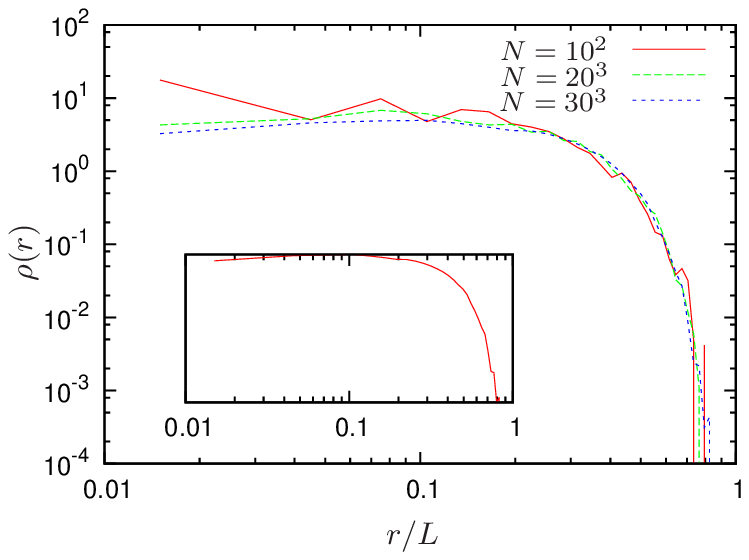}}}
\end{center}
\caption{{Results of simulations for the case of gravity ($\gamma=1$):
(a) Evolution of the total kinetic energy normalized to its initial value, for $N=8000$ and 
different values of $\epsilon$,  i.e., the set of simulations $\mathcal C(1;8000,\epsilon/L)$; (b) 
evolution of the normalized total kinetic energy with $\epsilon=0.01$ and a range of different values
of $N$, i.e., the set of simulations $\mathcal C(1;N,0.01)$; (c) velocity distribution for the simulation 
$\mathcal C(1;8000,0.002)$ at $t=20\tau_{dyn}$ and  (d) evolution of $\phi_{4}$ and $\phi_{6}$ for the simulations $\mathcal C(1;8000,0.002)$ and $\mathcal C(1;8000,0.02)$ at $t=20\tau_{dyn}$, 
(e) density distribution  for the simulations $\mathcal C(1;8000,\epsilon)$ at varying $\epsilon$ 
and $t=20\tau_{dyn} $; (f) density distribution  for the simulations $\mathcal C(1;i,0.01)$ at 
$t=20\tau_{dyn}$ and (inset) the same quantity for the simulation $\mathcal C(1;30^3,0.01)$ in log-log scale (note the density drops rapidly at $R/L\approx 1/3$)}.}
\end{figure*}

In this section we check our numerical and  analytical results using the
canonical much studied case of gravity as an established benchmark. 

\subsection{Qualitative inspection of evolution}

Fig.~\ref{N1_pot_eps_some} shows the evolution of the total kinetic 
energy normalized  to its initial value at $t=0$, for different values 
of the softening $\epsilon$ simulated. 
We observe that, for sufficiently 
small softening, and sufficiently short times, the curves match very well: 
we interpret this to be because they are following the same
mean-field evolution. Further the kinetic energy (and viral ratio)
shows a rapid relaxation (by $t \approx \tau_{dyn}$) to relatively
small and progressively damped oscillations around an
approximately stationary value. This is the familiar mean-field
relaxation to a QSS, which in practice we will consider to
be established below from
$t \approx 10 \tau_{dyn}$.  For larger times
we observe a slow linear drift in time of the average value of the kinetic
energy, which can be interpreted as a signature of the slow
collisional relaxation process. As predicted by Eq.~\eqref{rate_gr-soft},
the collisional relaxation is suppressed increasing the softening.

Fig.~\ref{N1_pot_N} compares the evolution of systems with
a fixed (compact) softening but different number of particles.
We observe a similar behavior to that in the previous plot,
and very consistent with the interpretation given of this
evolution as the relaxation to a QSS:  we observe a 
drift away from the almost stationary kinetic energy
which  develops more slowly as the number of particles
$N$ increases.  

Fig.~\ref{N1_L1_vel_425_ep0.001} shows, for the simulation
$\mathcal C(1;8000,0.002)$, the velocity distribution at $t=20 \tau_{dyn}$.
We observe that the tails of the distribution are clearly
non-Gaussian, and thus that the system is not
at thermal equilibrium. This is confirmed by the evolution 
of the functions $\phi_4$ and $\phi_6$, which are plotted 
in Fig.~\ref{phi_N1_L1}.  They are clearly non-zero, indicating
a non-Gaussian state, and further, show manifestly a slow
growth on a longer time-scale which is indicative of an
evolution towards a thermal state.  Finally, as shown 
in Fig.~\ref{N1_rho_eps} and \ref{N1_rho_N} respectively,
the density profile (i.e. mean density in spherical shells
centered on the center of mass of the system) at
$t=20\tau_{dyn}$ 
are substantially independent of the parameters 
$\epsilon$ and $N$, as they should be if this profile
is characteristic of a QSS.

\subsection{Scaling of the relaxation rate}
\label{scaling-grav}
\begin{figure*}
 \begin{center}
\subfloat[]{\label{N1_delta_eps_some}\includegraphics*[width=8cm]{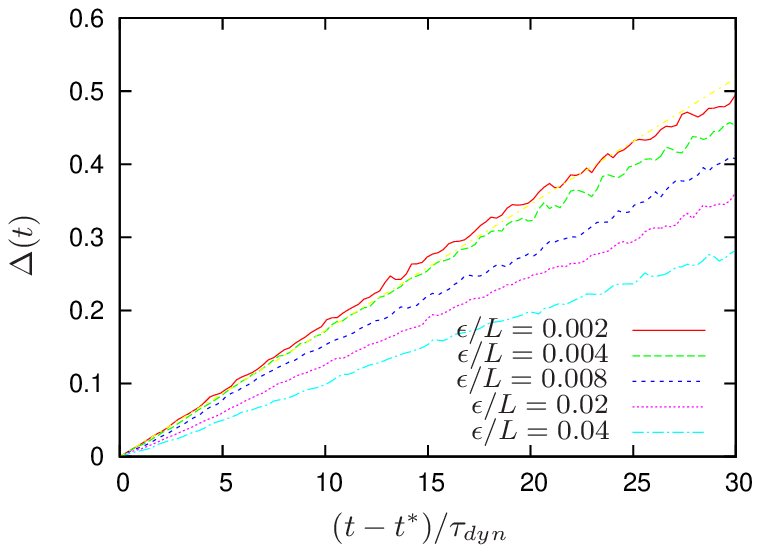}}
\subfloat[]{\label{N1_delta_e_N}\includegraphics*[width=8cm]{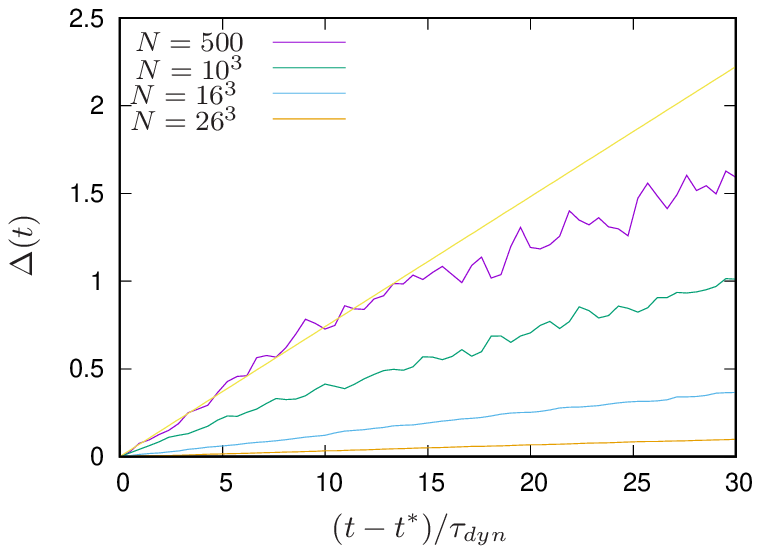}}\\
\subfloat[]{\label{slope_N1_eps}\includegraphics*[width=8cm]{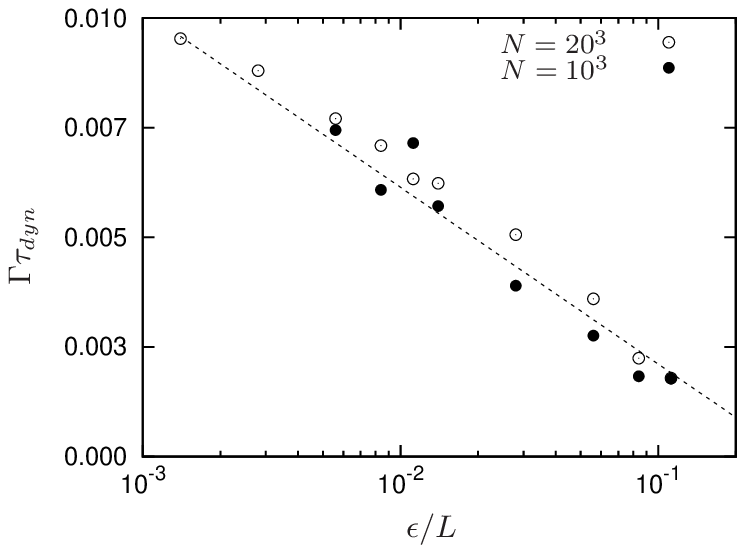}}
\subfloat[]{\label{slope_N1_N}\includegraphics*[width=8cm]{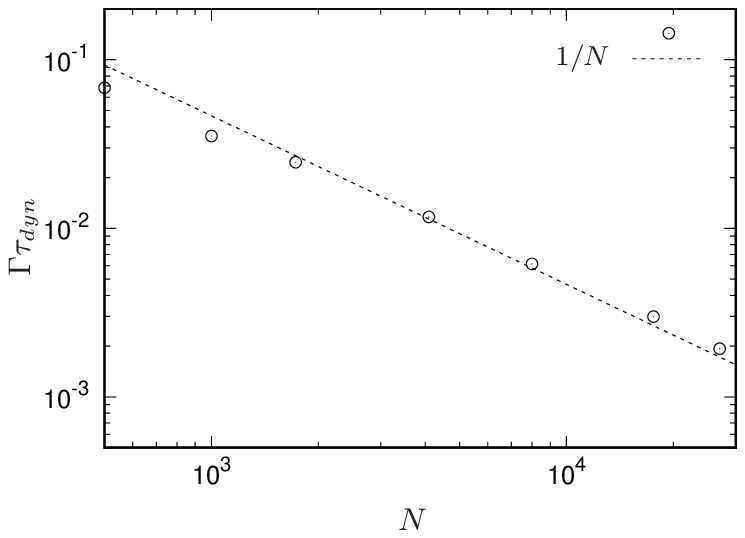}}
\caption{Measures of relaxation times for the case of gravity ($\gamma=1$):
(a) Evolution of the indicator $\Delta(t)$ for chosen values of $\epsilon$ and fixed $N=8000$, i.e., 
in the set of simulations  $\mathcal C(1;8000,\epsilon/L)$; (b) evolution of  $\Delta(t)$ for the range
of different $N$ simulated and $\epsilon=0.01$ i.e., the set of simulations $\mathcal C(1;N,0.01)$; (c) 
plot of $\Gamma \tau_{dyn}$ as a function of $\epsilon/L$ for both $N=8000$ and $N=1000$. In the latter case, following Eq.~\eqref{rate_gr-soft},  the amplitude of the relaxation rate has been multiplied by 
a factor $8$ in order to collapse both the scalings on a single curve; the straight line is the theoretical scaling $\Gamma\tau_{dyn}\sim \epsilon^{-1}$; (d) plot of $\Gamma$ as a function of $N$ for fixed
$\epsilon/L=0.01$.}
\end{center}
\end{figure*}
Figs.~\ref{N1_delta_eps_some} and \ref{N1_delta_e_N} show the
evolution of the collisional relaxation parameter $\Delta(t)$, defined
in Eq.\eqref{estimate_micro}, as a function of time, for different
values of $\epsilon$ and $N$. We estimate the relaxation rate as the
slope of a linear fit to $\Delta(t)$ at short times.  Inspecting
Fig.~\ref{N1_pot_eps_some} or \ref{N1_pot_N}, we assume that the QSS
has been reached at $t=10\tau_{dyn}$, and we take the reference time
$t^*$ to evaluate the slope of $\Delta(t)$ as $t^*=20\tau_{dyn}$. We
can estimate the value of $b_0$ using Eq.~\eqref{b0gamma} by measuring
the relative velocity from the simulation. This gives $b_0/L\approx
8.8\times 10^{-5}$. As this is considerably smaller even that the
smallest softening used, we expect that the relaxation rate will scale
as in Eq.~\eqref{rate_gr-soft} rather than Eq.~\eqref{rate_gr}.
We show in Figs.~\ref{slope_N1_eps} and \ref{slope_N1_N} the measured scalings of the relaxation rate with $\epsilon$ and $N$ respectively. We observe that there is 
indeed very good agreement with the theoretical scaling 
of Eq.~\eqref{rate_gr-soft}.

\section{Results for potentials with $\gamma \ne 1$}
\label{results-gne1}

We now consider the case of power law interactions other than
gravity. We consider first pair interactions which decrease more
rapidly at large separations than the gravitational one, i.e., 
$\gamma > 1$, and then the case $\gamma < 1$.

\subsection{Interactions decaying faster than gravity ($\gamma >1$)}

We present results for two specific cases: $\gamma = 5/4$ and
$\gamma = 3/2$. As discussed above we do not consider even larger 
values because, as predicted by the our analytical calculations, 
the two body collision rates indeed increase rapidly as $\gamma$ does, 
making it more and more difficult numerically to separate the associated 
time scale from the  mean field one. Indeed from Eq.~ \eqref{rate_slr-soft}
it follows that, at fixed $N$, the relaxation rate scales 
as $\epsilon^{-2\gamma}$. 

\begin{figure*}
\begin{center}
\subfloat[]{\label{N5d4_pot_eps_some}\includegraphics*[width=8cm]{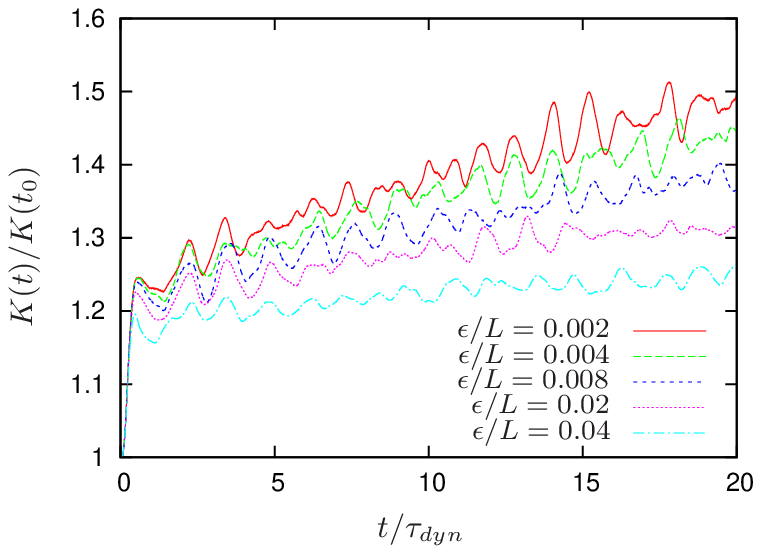}}
\subfloat[]{\label{N5d4_pot_N}\includegraphics*[width=8cm]{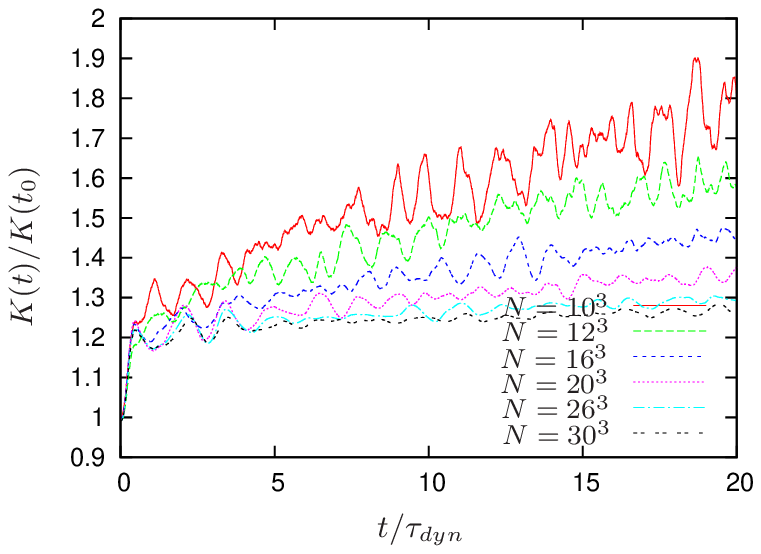}}\\
\subfloat[]{\label{N_L1_vel_210_ep0.001}\includegraphics*[width=8cm]{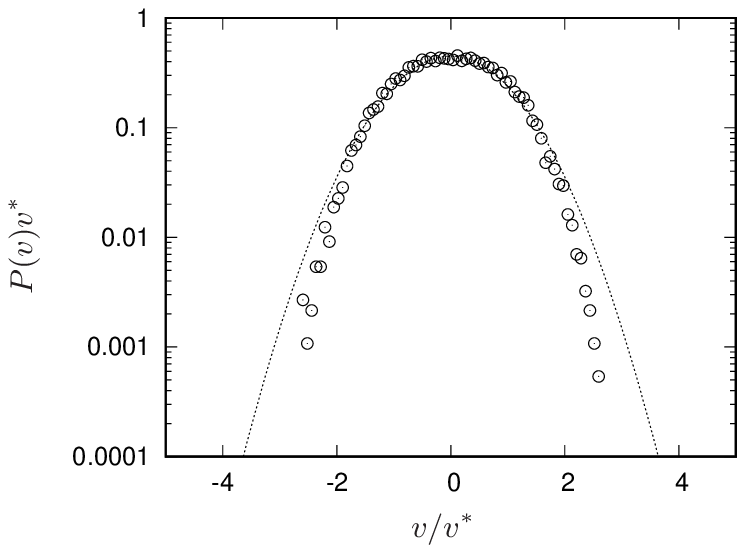}}
\subfloat[]{\label{phi_L1}\includegraphics*[width=8cm]{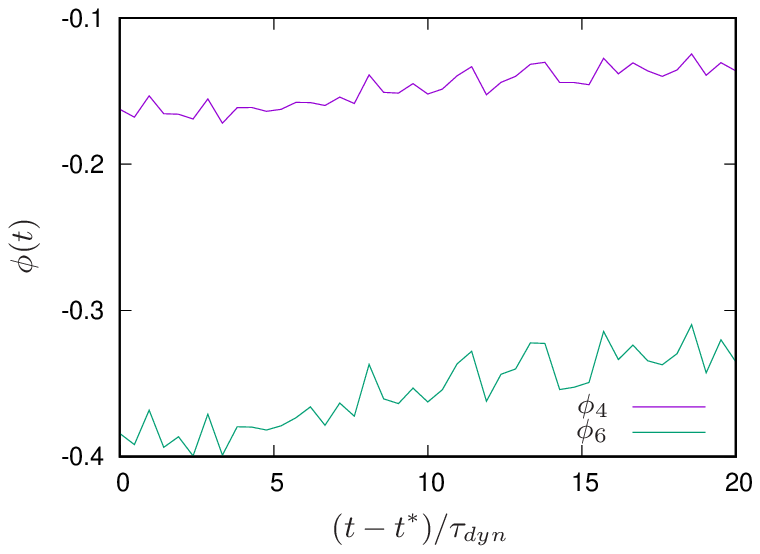}}
\caption{Results of simulations for the case $\gamma=5/4$:
(a) Evolution of the normalized total kinetic energy for different values of $\epsilon$ at fixed $N=8000$, 
i.e., the set of simulations $\mathcal C(5/4;8000,\epsilon)$, (b) same quantity but for varying $N$ and 
fixed $\epsilon/L$, (c) velocity distribution for the simulation $\mathcal C(5/4;8000,0.002)$ at $t=10 \tau_{dyn}$, 
and (d) evolution of $\phi_{4}$ and $\phi_{6}$ for the simulation 
$\mathcal C(5/4;8000,0.002)$.}
\end{center}
\end{figure*}

Figs.~\ref{N5d4_pot_eps_some} and \ref{N5d4_pot_N} 
display results for the case $\gamma=5/4$, in a
manner completely analogous to the case of gravity
above. We observe a very similar behavior to that in the 
gravitational case: the curves of the total kinetic energy 
are superimposed at the early stage of evolution, 
and start to separate as time increases. Consistent
with the interpretation of this drift as due to two
body relaxation, we observe that it becomes 
slower for larger $N$ and larger $\epsilon$.

For our quantitative analysis of the collisional
relaxation we choose the reference time 
$t^*=10\tau_{dyn}$, as the oscillations 
about the QSS are small 
by this time. For the case $\gamma = 3/2$,
for which we do not show the data (which is
qualitatively very similar), we take
$t^*=5\tau_{dyn}$.  Fig.~\ref{N5d4_pot_eps_some} 
show the time evolution, for $t> t^*$,  of $\Delta(t)$,
$N=8000$ and different values of the softening $\epsilon$.
The velocity distribution at 
$t=2t^*$ is plotted 
in Fig.~\ref{N_L1_vel_210_ep0.001},
and the evolution of the parameters $\phi_4$ and $\phi_6$
as a function of time, shows that the system has
a velocity distribution quite close to Gaussian, and
apparently evolves progressively closer to such
a distribution, as expected.  Very similar behaviors
are observed for the case $\gamma=3/2$. 
We do not plot the radial density profile, but it has
a form which varies little with $\gamma$ and
thus very similar to that plotted in Fig.~\ref{N1_rho_eps}.

We estimate the relaxation rate in the same manner as we
did above for the case of gravity, using the evolution of the 
indicator $\Delta(t)$ (which we do not plot).
Estimating again the value of $b_0$ using Eq.~\eqref{b0gamma}, 
we obtain $b_0/L\approx 2.5\times 10^{-4}$ for $\gamma =5/4$, and
$b_0/L\approx 7.3\times 10^{-4}$  for $\gamma=3/2$.
As in the case of gravity, these are therefore much smaller than the minimal 
softening $\epsilon$ used, and we thus expect that the scaling of the 
relaxation rate should be given by Eq.~\eqref{rate_slr-soft}.

\begin{figure*}
\begin{center}
\subfloat[]{\label{slope_eps}\includegraphics*[width=8cm]{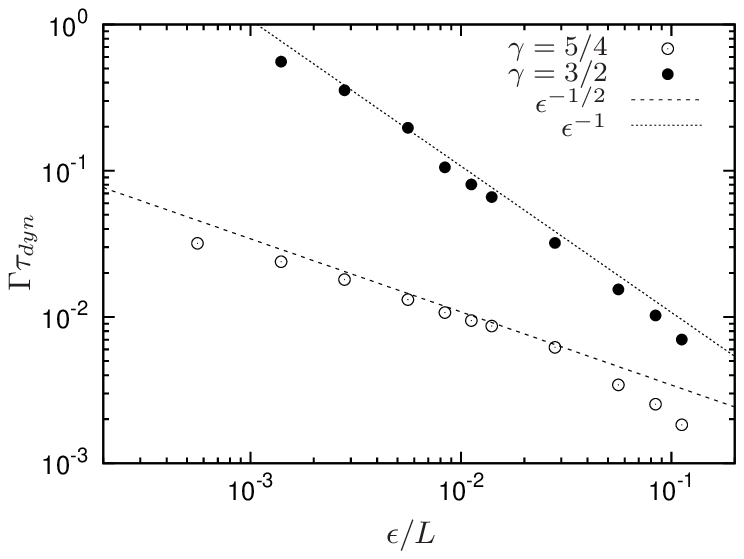}}
\subfloat[]{\label{slope_N}\includegraphics*[width=8cm]{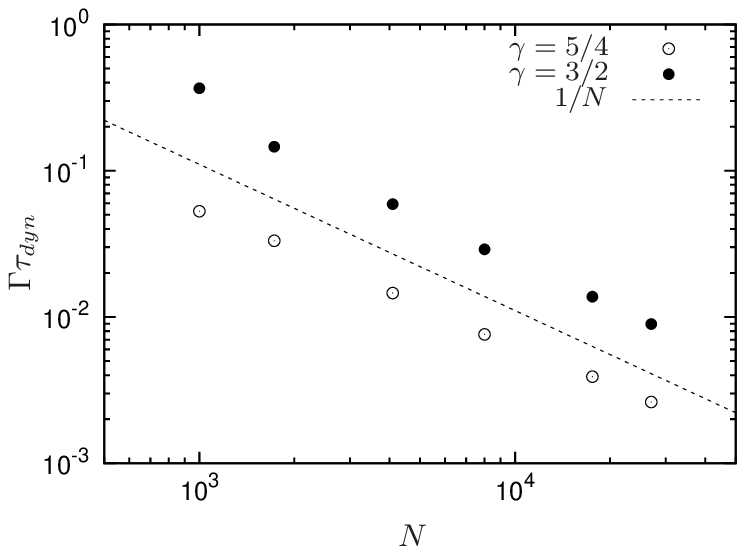}}\\
\subfloat[]{\label{N5d4_kin_res}\includegraphics*[width=8cm]{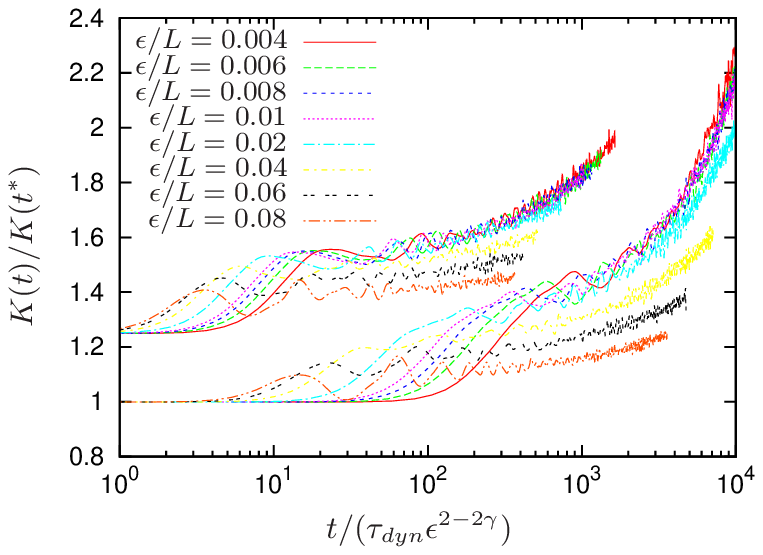}}
\subfloat[]{\label{collapse_macro_N5d4}\includegraphics*[width=8cm]{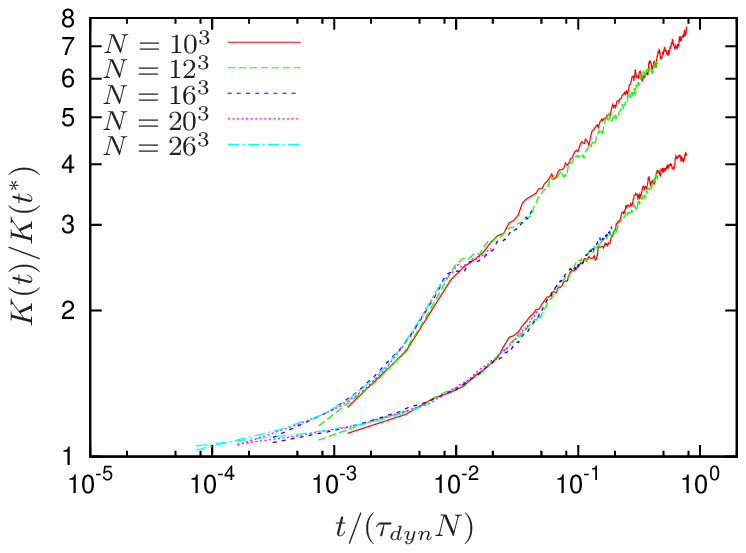}}
\caption{Tests of scaling of measured relaxation rates: 
(a) $\Gamma \tau_{dyn}$ as a function of $\epsilon$ (compact softening), for the cases $\gamma=5/4$ and $\gamma=3/2$ in simulations, and (b) as a function of $N$ for $\gamma=5/4$ and $\gamma=3/2$, (c) collapse plot  at $N=20^3$ constant and varying $\epsilon$ for $\gamma =5/4$  (upper curves, all the curves have been multiplied by a factor of $1.25$) and $\gamma=3/2$ and (d) collapse plot at constant $\epsilon/L=0.1$ and varying $N$ for $\gamma=5/4$ (lower plot) and $\gamma=3/2$.}
\end{center}
\end{figure*}

Fig.~\ref{slope_eps} shows the measured relaxation rate for a range
of softenings $\epsilon$ (for compact softening) at constant
particle number $N=8000$,  for both  $\gamma= 5/4$ and $\gamma = 3/2$.
Fig.~\ref{slope_N} shows the scaling of the relaxation rate at varying $N$
and constant $\epsilon/L = 0.01$. The error bars
have been determined as the statistical error in the fit of $\Delta$, 
and are smaller than the size of the symbols. We observe that 
there is very good agreement between the scalings
measured and the theoretical one \eqref{rate_slr-soft}. For the
largest values of $\epsilon$ we observe a departure from the
theoretical scaling. This is due to the finite size of the system
(when $\epsilon$ is around one tenth of the size of the system,
where the latter is estimated from the fall-off of the density profile).

\subsection{Relaxation at longer times}

In the previous subsections we have considered collisional
relaxation over time scales over which the parameters
used to monitor evolution change by a small amount. 
In principle the predicted scalings should apply also 
on longer time scales, provided  the scale introduced by the 
softening length is sufficiently small that it does not affect 
significantly the properties of the QSS. 

Fig.~\ref{N5d4_kin_res} shows the normalized total kinetic energy for
the case $\gamma =5/4$ (top curves) and $\gamma = 3/2$ (bottom curves)
for a constant particle number $N$ and a range of $\epsilon$. The time
axis has been rescaled following the theoretical scaling
\eqref{rate_slr}.  We observe a good superposition of the curves for
the smaller values of $\epsilon$, while for softening approaching the
size of the system the observed relaxation rate is suppressed compared
to the theoretical prediction, just as for the shorter time relaxation
(see Fig.~\ref{slope_eps}).  Fig.~\ref{collapse_macro_N5d4} shows an
analogous collapse plot but for a (small) constant $\epsilon$ and
varying $N$, with the time axis now rescaled with $N$ following
\eqref{rate_slr}.  We observe a very good matching between the
different curves over the whole duration of the runs.

\subsection{Results: case $\gamma <1$}

\begin{figure}
\begin{center}
\subfloat[]{\label{N1d2_delta_eps}{\includegraphics*[width=8cm]{{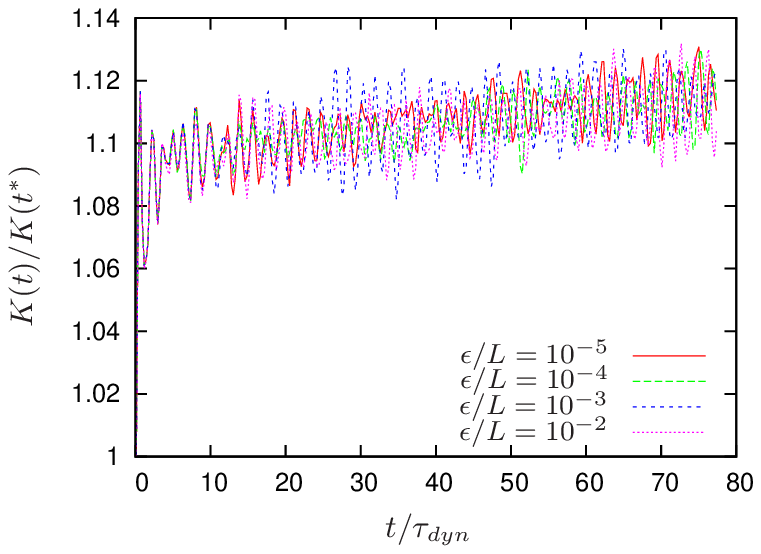}}}}\\
\subfloat[]{\label{N1d2_slope}\includegraphics*[width=8cm]{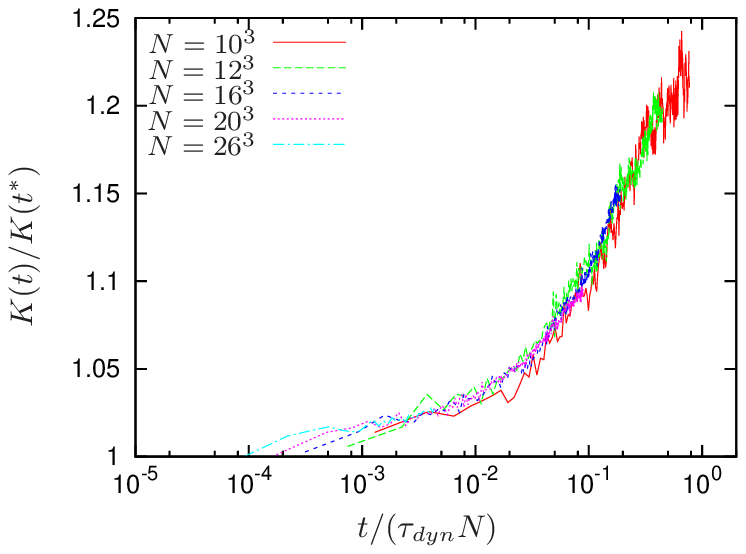}}
\caption{Evolution of the kinetic energy for systems with $\gamma = 1/2$: (a) for a range of 
different values of $\epsilon$ at fixed  $N=8000$, and (b)  for a range of $N$ 
different number of particles at fixed $\epsilon$=0.0028 . In the latter plot the time
variable has been rescaled with $N$ in line with the theoretically predicted scaling of 
Eq.}
\end{center}
\end{figure}

In this case we have seen that the scaling of the relaxation rate 
is very simple: inversely proportional to $N$, and independent of 
the softening (cf. Eq.~\eqref{rate_llr-soft}). This behavior is a 
consequence of the fact that the dominant contribution comes 
from the largest impact factor, which we have assumed to scale 
with the system size. To test this prediction we have simulated 
the case $\gamma=1/2$. Fig.~\ref{N1d2_delta_eps}  shows the 
evolution of the normalized kinetic energy as function of time 
for a range of (compact) softenings $\epsilon$, while 
Fig.~\ref{N1d2_slope} shows the same quantity for
a range of $N$ at fixed (small) $\epsilon$, as a function of a 
time variable linearly rescaled with $N$ in accordance with
the the predicted scaling. We observe that the results are
in excellent agreement with the theoretical predictions.

\section{Tests of analytical predictions: beyond scaling}
\label{quantitative}

In the previous sections we have tested numerically 
the validity of the theoretical scaling relations derived in the
first part of the paper. We now examine further how well
the amplitudes of the measured relaxation rates match
the predictions.  

As we have discussed (see also \cite{marcos_13b}), the approach we have adopted
in deriving two body collision rates, following that used  originally by Chandrasekhar for 
gravity, makes a number of very strong simplifying assumptions which make the
calculation intrinsically inaccurate, notably: spatial homogeneity of the system and
the assumption that all collisions take place at fixed relative velocity fixed by the
global velocity dispersion.  Further the ``largest impact factor", which we  taken it to be 
given by the system size, is not in fact a precisely defined quantity and indeed
it is often treated as a free parameter (see e.g. \cite{merritt_book} for a discussion 
in the context of the  orbit-averaging technique).  Other collisional effects which have
been identified through the study of kinetic equations, such as orbit resonances 
and various collective effects (see e.g.~\cite{chavanis_12b}), are also evidently not 
taken into account. Thus, even if incoherent two body scatterings are the 
dominant collisional process, we cannot expect the calculation method
given to provide a precise prediction for the relaxation rates. Nevertheless
the fact that the predicted scalings turn out to be in such good agreement
with those observed, one would expect the quantitative discrepancies
not to be too large.  

\subsection{Effect of softening function}

\begin{figure*}
\begin{center}
\subfloat[]{\label{fig-plummer_5/4}\includegraphics*[width=8cm]{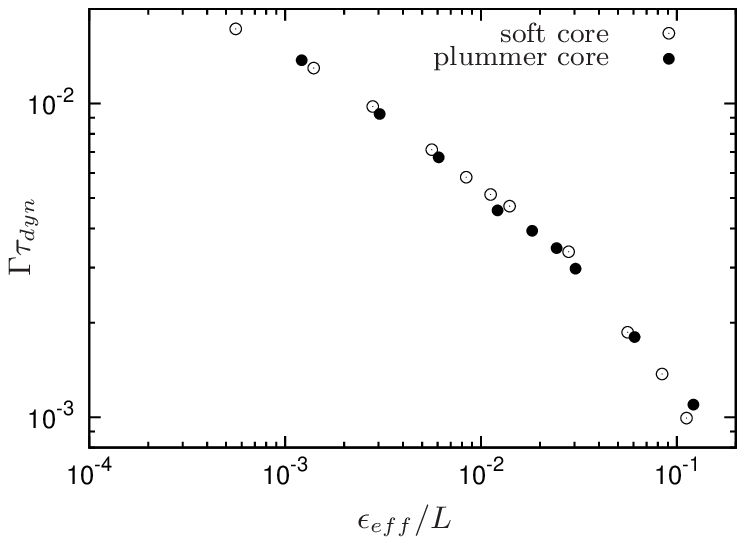}}
\subfloat[]{\label{fig-plummer_3/2}\includegraphics*[width=8cm]{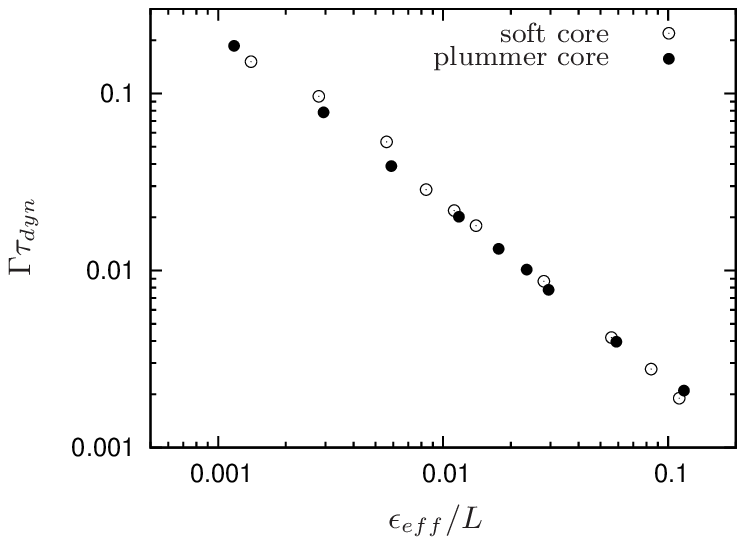}}
\caption{Measured relaxation rates as a function of $\epsilon_{eff}$ 
 for the two different softening functions,
for (a) $\gamma=5/4$ and (b) $\gamma=3/2$.}
\end{center}
\end{figure*}

In subsections \ref{theo-soft} and \ref{softenings} we have discussed 
how the softening of the potential at small scales effects the
predicted relaxation rate. The predicted modification depends, in 
general,  not just on the value of the softening scale, but on the 
detailed form of the softened potential. We have noted, however,
that for $\epsilon \gg b_0$, the effect of any such smoothing is
an overall amplitude shift (cf. Fig.~(\ref{rutherford_softening3})).
This allowed us to define, for any softening potential, a
constant $\alpha$ giving an effective softening 
$\epsilon_{eff}=\alpha \epsilon$. The latter is the value
of the softening of a reference softened potential which is 
sharply cut-off at $\epsilon_{eff}$,  which gives the same
predicted relaxation rate as the actual softened potential.
The values of $\alpha$ for the two potentials (compact and
Plummer) we have employed are given in Table \ref{table-soft}.

Thus the theoretical calculations of the two body relaxation rates
make a prediction about the relative amplitude of the relaxation 
rates for our two different smoothings, which we should expect
to hold even if the prediction of the absolute amplitude of both
is (expected to be) incorrect.  Figs.~\ref{fig-plummer_5/4} and \ref{fig-plummer_3/2} 
shows the relaxation rate measured in simulations with $N=8000$,
as a function of the calculated $\epsilon_{eff}$ over a wide range. 
The superposition of the two curves is almost perfect, in line
with the theoretical prediction.

\subsection{Detailed comparison of relaxation rates}
We now compare directly the amplitudes of the predicted and measured
relaxation rates.  Tab.~\ref{table-comp} shows, for the different
values of $\gamma$ we have simulated, the results of this
comparison. The second column gives the numerical value of $b_0\approx
\left(g/( m\langle v^2\rangle)\right)^{1/\gamma}$, where $\langle
v^2\rangle $ is the velocity dispersion measured at $t=20\tau_{dyn}$
in the simulations (we have used that $\langle V^2\rangle \simeq 2
\langle v^2\rangle$). Using this value for $b_0$, and taking $R=0.3$
for the system size (cf. Figs.~\ref{N1_rho_eps} and \ref{N1_rho_N}),
we have calculated numerically the predicted $\Gamma \tau_{\rm dyn}$
shown in the third column (``Theory'') using
Eq.~\eqref{total-change-v}.  The fourth column (``Numerics") gives the
value of $\Gamma \tau_{\rm dyn}$ estimated in our simulations from the
short time evolution of the normalized total kinetic energy
$K(t)/K(t_0)$ as described in Sect.~\ref{scaling-grav}. Comparing the
last columns we find that, despite the many crude approximations
performed in the derivation of the relaxation rate we obtain, as we
have seen, not only the right scaling with the relevant parameters,
but also a relatively good quantitative agreement for the amplitudes
for all the cases simulated, with an overall discrepancy in the
normalization varying between a factor one and eight. Moreover, we
observe that, as $\gamma$ increases, the agreement is better.  This is
compatible with the idea that as the interaction becomes less long
range, the resonances between particles with different frequencies
become less important and a local approximation better and better (see
e.g. \cite{heyvaerts_10}).
\begin{table}
\begin{tabular}{|c|c|c|c|c|}
\hline
$\gamma$  &  $b_0\approx (g/(2 m\langle V^2\rangle))^{1/\gamma}$ & Theory & Numerics\\
 \hline \hline
1/2 &  $9.2\times 10^{-8}$ & $7.4 \times 10^{-3}$ & $4.6 \times 10^{-4}$ \\
3/4 &  $8.4\times 10^{-6}$ & $1.4\times 10^{-2}$ & $1.1 \times 10^{-3}$ \\
1   &  $8.8\times 10^{-5}$& $0.016$ & $4.6 \times 10^{-3}$\\
5/4 &  $3.7 \times 10^{-4}$ & $0.059$ & $0.023$ \\
3/2 &  $8.7\times10^{-4}$ & $0.017$ & $0.24$\\
\hline
\end{tabular}
\caption{Comparison of the theoretical and  measured relaxation rates in the simulations. The second column corresponds to an estimation of $b_0$, the third one to the estimation of $\langle|\Delta \bV^2|\rangle/|V^2|$ using Eq.~\eqref{total-change-v} and the fourth one the relaxation time measured in the simulations (see text).}
\label{table-comp}
\end{table}

\subsection{Constraining the maximum impact factor}

Going back to the original derivation of the two body relaxation rate 
by Chandrasekhar there has been a debate about the correct
choice of the {\it{maximum}} impact parameter. In section 
\ref{chandrasekhar} we have argued that it should be
assumed to be of the order of the size of the system, and 
we have obtained our results making this hypothesis. 

For the case $\gamma\le (d-1)/2$, which is dominated by the largest 
impact factors, we can in principle test this hypothesis.
If, instead of Eq.~\eqref{rate_llr}, we fix an arbitrary maximum 
parameter $b_{max}$, it is straightforward to show that we obtain 
\be
\label{result_bmax}
\Gamma\tau_{dyn} =   \tilde C N^{-1}\left(\frac{R}{b_{max}}\right)^{2\gamma-d+1},
\ee
where $\tilde C$ is a numerical coefficient (depending only on 
$\gamma$ and $d$). 
If we now assume that $b_{max}\sim RN^{-\alpha}$ 
we obtain 
\be
\label{result_bmax-mod}
\Gamma\tau_{dyn} \sim  N^{\beta},
\ee
where $\beta=\alpha(2\gamma-d+1)-1$. The case $\alpha=0$ corresponds to
the assumption we have made up to now, and the result \eqref{rate_llr}. 
The case $\alpha=1/d$ corresponds, on the other hand, to the assumption 
that $b_{max}$ scales in proportion to the inter-particle distance
(as originally assumed by Chandrasekhar \cite{chandra_42}). 
Now we have seen in section \ref{results-gne1} that the
scaling of the relaxation rate for the simulated cases $\gamma = 1/2$
and $\gamma=3/4$ are $\Gamma\tau_{dyn}\sim N^{-1}$, which 
are in agreement with $\beta=0$ and hence $b_{max}\sim R$. 

In the specific case $\gamma = (d-1)/2 $, i.e. gravity in $d=3$, it is
in fact possible  to {\it quantify} the maximum impact factor rather 
than just its scaling. Instead of Eq.~\eqref{rate_gr-soft} 
(replacing  $\epsilon$ by $\epsilon_{eff}$ following the discussion 
in Sect.~\ref{softenings}) we have 
\be
\label{result_bmax_grav}
\Gamma\tau_{dyn} = \tilde D N^{-1}\ln\left(\frac{b_{max}}{\epsilon_{eff}}\right),
\ee
where $\tilde D$ is a (calculable) numerical coefficient. 
Using the simulations presented in Sect.~\ref{simu-grav}, we
can fit very well the relaxation rate with
\be
\label{scaling_grav_eps}
\Gamma\tau_{dyn} =  \ln\left(\frac{L}{3\epsilon_{eff}}\right)\frac{7.2}{N}.
\ee
Comparing these last two equations, we have that $\alpha\approx 0$,
and, further, that $b_{max} \approx L/3\approx R/3$.  This size
corresponds with the sharp fall-off of the density profile shown in
the inset of Fig.~\ref{N1_rho_N}.  To check that $b_{max}$ does not
depend on $N$, we did another set of simulations with the same
parameters but $N=1000$ particles.  From these we obtained the scaling
of the relaxation rate as a function of $\epsilon$ plotted in
Fig.~\ref{slope_N1_eps}, in which, according to
Eq.~\eqref{rate_gr-soft}, the relaxation rate has been multiplied by a
factor of eight. We thus obtain very good agreement with the predicted
scaling.  Our findings confirm therefore the results of Farouki \&
Salpeter \cite{farouki_82,farouki_94}, who found that the maximum
impact parameter should be taken of order of the size of the system.

\section{Conclusion}

In this paper, we have studied collisional relaxation in systems
of particles interacting with a power-law potential
$v(r\to\infty) \sim 1/r^\gamma$ \eqref{pot}, introducing a 
regularization of the singularity in the force as $r \rightarrow 0$
when necessary.  In our analytical calculations we have generalized 
the ``Chandrasekhar approach" in the case of gravity to such
potentials. We have also  included the contribution of hard collisions
rather than just weak collisions, in which the mean field trajectories of the 
particles are weakly perturbed, which is the approximation usually
found in the literature, see e.g. \cite{chavanis_12b}. 
We have found that the collisional dynamics is dominated 
by 
\begin{itemize}

\item weak collisions, if $\gamma < (d-1)/2$, and

\item hard collisions, if $\gamma > (d-1)/2$, 

\end{itemize}
while the case $\gamma =(d-1)/2$, which corresponds to gravity in $d=3$,
is at the threshold.  Moreover we considered the large $N$, mean field (or Vlasov)
limit scaling of the two body relaxation rate, assuming the
considered particle system to be in viral equilibrium. 
{\it In absence of  force regularization} (other than an 
infinitesimal one  assumed implicitly to make two body collisions defined
for $\gamma >2$),  we found that this rate, expressed
in units of the characteristic time for mean-field dynamics
$\tau_{dyn}$,  vanishes in the large $N$ for $\gamma<d-1$,
and diverges in this limit for $\gamma>d-1$. This means 
that only in the former case does the mean-field 
limit of the dynamics exist for a virialized system; 
in the latter case it does not because the collisional 
relaxation completely dominates the mean field dynamics. 
Only in the former case, therefore, can a QSS be expected
to exist on a physically relevant time scale. 
This leads to the following {\it dynamical} classification of
interactions:
\begin{enumerate}
\item Power-law interactions is {\it dynamically} long-range if
  $\tau_{dyn}\ll\tau_{coll}$ for a sufficiently large number of
  particles, and in particular $\lim_{N\to\infty} \Gamma
  \tau_{dyn}=0$, which occurs for $\gamma<d-1$.  
\item The interaction is {\it dynamically} short range if
  $\tau_{dyn}\gg\tau_{coll}$ for a sufficiently large number of
  particles, and in particular $\lim_{N\to\infty} \Gamma
  \tau_{dyn}=\infty$, which occurs for $\gamma<d-1$.
\end{enumerate}
This classification was proposed initially \cite{gabrielli_10b}
on the basis of a formal analysis of convergence properties 
of the force on a particle in the thermodynamic limit, and 
subsequently in \cite{gabrielli_10} on the basis of the 
analysis detailed here. It has also been justified using 
different analytical approaches to the full kinetic theory 
of such systems \cite{chavanis_13, gabrielli_etal_2014}.
As noted in the introduction, this classification differs from
the usual one used to distinguish long range from short range 
interactions, according to the thermal equilibrium of the system, in
which the important feature is the integrability of the {\it
 potential}. There is therefore a range of $\gamma$, $d-1<\gamma <
d$, in which the interaction is {\it dynamically} short range, but
long-range according to its thermal equilibrium properties. In this
case, if the number of particles is sufficiently large, there
will be no QSS (as in short range systems), but the thermal
equilibrium state will present the typical features of a 
long-range system, i.e., spatial inhomogeneity,
inequivalence of ensembles etc.. 

We have also generalized these scalings when the inter--particle potential is regularized (``softened'') at small scales. With this regularization 
the case $\gamma\ge 2$  (in which the potential barrier cannot prevent the particles to collide for pure power--law potentials) becomes well defined.  In this case, the relaxation rate depends on the value of the softening length $\epsilon$ for interactions in which small impact factors play a predominant role, i.e., $\gamma\le (d-1)/2$.

We have presented, for $d=3$, detailed numerical results 
which support our theoretical findings. We have confirmed 
previous results in the literature for the gravitational case $\gamma=1$,
notably  for the scaling relations satisfied by the relaxation rate as function of the
softening $\epsilon$ and the number of particles $N$. Furthermore, using the 
scaling of the relaxation rate with $\epsilon$, we have found  very 
strong numerical evidence that the maximum impact parameter is related with 
the size of the system and not microscopic scales such as the
inter-particle distance. We have simulated also dynamically long-range 
cases $\gamma=5/4$ and $\gamma=3/2$, in which the collisional relaxation 
is dominated by collisions around the  minimum impact parameter,  obtaining 
again very good agreement with the theoretical scalings. 
For dynamically long-range systems dominated in our calculations
by collisions with the largest impact parameter, we have found find, as 
predicted, that a softening  in the potential does not affect the 
relaxation rate.

The natural extension of this work is the numerical study of
collisional relaxation allowing strong collisions, in order to check
the scalings of this regime derived in this paper. For such study, it
is necessary to develop very refined integration schemes in order to
integrate properly such collisions. Another interesting perspective is
to study the problem with a more rigorous approach using the
angle-action variables (with probably also many approximations because
it is a very complicated formalism) in order to describe more
precisely the relaxation dynamics, and in particular study more precisely 
the validity of the  Chandrasekhar approximation as a function of the range of the interaction 
$\gamma$.

\section*{Acknowledgments}

We acknowledge many useful discussions with J. Morand, F. Sicard and
P. Viot.  This work was partly supported by the ANR 09-JCJC-009401
INTERLOP project and the CNPq (National Council for Scientific
Development, Brazil). Numerical simulations have been performed at the
cluster of the SIGAMM hosted at ``Observatoire de C\^ote d'Azur'',
Universit\'e de Nice -- Sophia Antipolis and the HPC and visualization
resources of the Centre de Calcul Interactif hosted by Universit\'e
Nice Sophia Antipolis.

\appendix

\section{An alternative derivation of the change in perpendicular velocity due to a collision}
\label{alternative-derivation}
It is interesting to derive Eq.~\eqref{change-v-perp-soft} simpler method which can give more physical insight. We can compute
the change in perpendicular velocity integrating the perpendicular
component of the force for all the duration of the collision, assuming
that the relative trajectories is unperturbed with constant relative
velocity $V$:
\be
F_\perp =   \frac{\gamma g}{b^{\gamma+1}}\left[1+\left(\frac{Vt}{b}\right)^2\right]^{-(\frac{\gamma}{2}+1)},
\ee
The change in the perpendicular component of the velocity in a time $2t_c $ is thus
\bea
\label{integral0}
|\Delta \bV_\perp| 
&=&\frac{\gamma g}{m b^{\gamma+1}} \int_{-t_c}^{t_c} dt \left[1+\left(\frac{Vt}{b}\right)^2\right]^{-(\frac{\gamma}{2}+1)}\\
&=& \frac{\gamma g}{m b^\gamma V}\int_{\frac{Vt_c}{b}}^{-\frac{Vt_c}{b}}  ds (1+s^2)^{-(\frac{\gamma}{2}+1)} \\
&\simeq&  \gamma\left(\frac{b_0}{b}\right)^\gamma\int_{-\infty}^{\infty}  ds (1+s^2)^{-(\frac{\gamma}{2}+1)};
\label{deflection}
\eea
Taking the limit $t_c\to\infty$ and performing the integral we obtain exactly \eqref{change-v-perp-soft}. 

\section{Exact form of the potential with a soft core}
\label{appendix_potential}

The potential $v(r,\epsilon)$ is, for $r\ge\epsilon$, exactly 
\be
v(r\ge\epsilon,\epsilon)=\frac{g}{r^\gamma}.
\ee
We define $u=r/\epsilon$. For $u<1$ we use the following form of the potential for \emph{soft} core softenings:
\begin{itemize}
\item $\gamma = 1/2$:
\be
v(u,1)\epsilon^{1/2}=15.75 u^2 - 22.5u^3 + 8.75u^4,
\ee
\item $\gamma = 3/4$:
\be
v(u,1)\epsilon^{3/4}= 11.875 u^2 -17.4167  u^3 + 6.875  u^4,
\ee
\item $\gamma = 1$:
\be
v(u,1)\epsilon= 10 u^2 - 15  u^3 + 6 u^4,
\ee
\item $\gamma = 5/4$:
\be
v(u,1)\epsilon^{5/4}= 8.925 u^2  - 13.65u^3 + 5.525 u^4,
\ee
\item $\gamma = 3/2$:
\be
v(u,1)\epsilon^{3/2}= 8.25 u^2 - 12.8333333 u^3 + 5.25 u^4.
\ee
\begin{figure}
\begin{center}
{\includegraphics*[width=8cm]{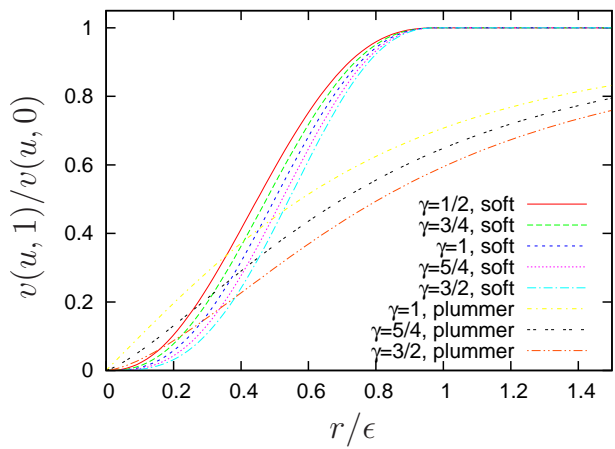}}
\caption{Softened potentials used in the paper normalized to the unsoftened one.}
\label{fig1}
\end{center}
\end{figure}
\end{itemize}

\end{document}